\documentclass{aa} 

\usepackage[colorlinks=true, linkcolor=blue, citecolor=blue, urlcolor=blue]{hyperref}
\makeatletter
\renewcommand*\aa@pageof{, page \thepage{} of \pageref*{LastPage}}
\makeatother
\usepackage{graphicx}
\usepackage{txfonts}
\usepackage{siunitx}
\usepackage{booktabs}
\newcommand{\Vaisala}{V\"{a}is\"{a}l\"{a}}
\usepackage{ulem}

\begin{document}

   \title{Tayler-Spruit dynamo in binary neutron star merger remnants}

   \author{Alexis Reboul-Salze
          \inst{1}, Paul Barr\`{e}re\inst{2}, Kenta Kiuchi \inst{1,5},  J\'{e}r\^{o}me Guilet \inst{3}, Rapha\"{e}l Raynaud \inst{4}, Sho Fujibayashi \inst{6,7,1}  \and Masaru Shibata\inst{1,5}}

   \institute{\inst{1} Max Planck Institute for Gravitational Physics (Albert Einstein Institute), D-14476 Potsdam, Germany\\
              \email{alexis.reboul-salze@aei.mpg.de}  \\
              \inst{2}{Observatoire de Genève, Université de Genève, 51 Ch. Pegasi, 1290 Versoix, Switzerland}\\
              \inst{3}Universit\'e Paris-Saclay, Universit\'e Paris Cit\'e, CEA, CNRS, AIM, 91191, Gif-sur-Yvette, France\\
              \inst{4}Universit\'e Paris Cit\'e, Universit\'e Paris-Saclay, CNRS, CEA, AIM, F-91191 Gif-sur-Yvette, France\\
              \inst{5} Center for Gravitational Physics and Quantum Information, Yukawa Institute for Theoretical Physics, Kyoto University, Kyoto 606-8502, Japan\\
              \inst{6} Frontier Research Institute for Interdisciplinary Sciences, Tohoku University, Sendai 980-8578, Japan\\
              \inst{7} Astronomical Institute, Graduate School of Science, Tohoku University, Sendai 980-8578, Japan
             }

   \date{Received ??; ??}

 
  \abstract
   {In binary neutron star mergers, the remnant can be stabilized by differential rotation before it collapses into a black hole. Therefore, the angular momentum transport mechanisms are crucial for predicting the lifetime of the hypermassive neutron star. One such mechanism is the Tayler-Spruit dynamo, and recent simulations have shown that it could grow in proto-neutron stars formed during supernova explosions.}
   {We aim to investigate whether hypermassive neutron stars with high neutrino viscosity could be unstable to the Tayler-Spruit dynamo and study how magnetic fields would evolve in this context.}
   {Using a one-zone model based on the result of a 3D GRMHD simulation, we investigate the time evolution of the magnetic fields generated by the Tayler-Spruit dynamo. In addition, we analyze the dynamics of the 3D GRMHD simulation to determine whether the dynamo is present.}
   {Our one-zone model predicts that the Tayler-Spruit dynamo can increase the toroidal magnetic field to $ \ge  10^{17}$ G and the dipole field to amplitudes $\ge  10^{16}$ G. The dynamo's growth timescale depends on the initial large-scale magnetic field right after the merger. In the case of a long-lived hypermassive neutron star, an initial magnetic field of $\ge 10^{12}$ G would be enough for the magnetic field to be amplified in a few seconds.
   However, we show that the resolution of the current GRMHD simulations is insufficient to resolve the Tayler-Spruit dynamo due to high numerical dissipation at small scales.
   }
   {We find that the Tayler-Spruit dynamo could occur in hypermassive neutron stars and shorten their lifetime, which would have consequences on multi-messenger observations.}

   \titlerunning{TS dynamo in BNS remnant}
   \authorrunning{A. Reboul-Salze, P. Barrere, J. Guilet, K. Kiuchi, , R. Raynaud, S. Fujibayashi and M. Shibata}
   \maketitle
%

\section{Introduction}

    The association of GW170817 and GRB 170817A confirmed that binary neutron star (BNS) mergers are the progenitors of some short gamma-ray bursts \citep{2017AbbottGRBandGW}.
    However, the nature of GW170817's remnant remains uncertain, and various physical parameters can impact the lifetime of the merger remnant before its collapse into a black hole. 
    Depending on the hot and dense matter equation of state and the masses of the neutron stars, different fates await the remnant after the merger: if the remnant mass $M$ remains lower than the maximum mass of a nonrotating neutron star $M_{\rm TOV}$, it will stay indefinitely stable.
    Otherwise, the remnant can promptly collapse into a black hole, or it can form a \textit{hypermassive} neutron star (HMNS) \citep{2000ApJ...528L..29B}, i.e. a neutron star that is stabilized by differential rotation while having a mass higher than the maximum mass of a uniformly rotating cold neutron star. 
    An HMNS will either (i) collapse into a black hole when differential rotation is transported \citep{Duez:2005cj,  Shibata:2005mz},  
    or (ii) be stabilized by solid body rotation, increasing $M_{\rm TOV}$ by around $\sim 20 \%$ \citep{1994CookHMNSEOS,1996LasotaRotatingNS,2016BreuHMNSstab,2024MusolinoHMNS}. 
    In the latter case, the remnant is called a \textit{supramassive} neutron star \citep{1994ApJ...422..227C} that will collapse after spinning down
    by magnetic dipole radiation on a timescale ranging from minutes to hours, or even longer.
    In the following, we call the long-lived remnant case when the remnant survives for O(1) s and it often corresponds to the \textit{supramassive} neutron star case.

    After the merger of two neutron stars, the nature of the remaining object can significantly affect the electromagnetic counterparts due to various factors. In the case of an HMNS, these include powerful neutrino luminosity, strong magnetic fields ($\ge 10^{15}$ G), and rotational kinetic energy reservoirs of order $10^{53}$ erg. Several studies have examined the impact of these factors on post-merger events \citep[e.g.,][]{2008MetzgerSGRB,2012BucciantiniSGRB,2013GaoBNSAfterglow,2014MetzgerSMNSemission,2014GompertzSGRB,2022SarinKilonovaSMNS}. 
    Magnetic fields, along with fast rotation, play a critical role in neutron star mergers as they can extract rotational kinetic energy and 
    drive powerful relativistic outflows 
    \citep[e.g.,][]{Shibata:2021xmo,2023CombiJetsfromBNS,2024Kiuchi}. Therefore, understanding the evolution of the magnetic field and the rotation profile of the HMNS is crucial for interpreting future multi-messenger observations.

    During the first few milliseconds after the merger, the magnetic field of the HMNS is amplified by the Kelvin-Helmholtz instability due to the presence of the shear layer at the contact interface between the two neutron stars \citep{1999RasioReviewKH,2006PriceKHBNSScience,2014KiuchiKH,2015KiuchiKHBNS,2024AguileraMiretHMNS}. %
    This generates a strong, small-scale magnetic field. However, another mechanism is necessary to get a large-scale magnetic field capable of driving an outflow.
    Both the magnetorotational instability (MRI) and the Tayler-Spruit (TS) dynamo could possibly fulfill this role, but act in different regions of the HMNS.
    Indeed, the remnant from the BNS merger has a rotation profile with an angular velocity that increases with radius until 10 km and then decreases afterwards \citep{2006ShibataTaniguchiBNS}. 
    Therefore this rotation profile is unstable to the MRI at radii larger than 10 km. This instability has been studied in the long-lived remnant case and 
    it was indicated that it could drive a luminous relativistic outflow \citep{2023CombiJetsfromBNS, 2024Kiuchi}. 
    By contrast, the differential rotation is such that the flow is stable to the MRI at radii smaller than 10 km, and core regions could instead be unstable to a Tayler-instability driven dynamo \citep{1973TaylerInstability,2002spruitTS}

    The Tayler instability is a type of magnetic instability in flows of electrically conducting, stably stratified fluids where a strong azimuthal magnetic field can become unstable, causing disruptions due to the Lorentz forces that arise from the interaction of currents along the axis of symmetry and magnetic fields \citep{1973TaylerInstability,1985Pitts}.
     The first model of a dynamo driven by the Tayler instability was proposed by \citet{2002spruitTS} for stably stratified, differentially rotating regions. 
     \citet{2019FullerTaylerSpruit} have proposed an alternative description to address criticisms of Spruit's model \citep[see][]{2007DenissenkovTS,2007ZahnTS}. For a long time, numerical simulations were unable to capture the TS dynamo, but recent numerical simulations have finally provided numerical evidence that this dynamo exists, both in the presence of negative shear \citep{2023PetitdemangeTS,2023daniel,2024PetitdemangeTS} and positive shear \citep{2023BarrereTSsim,2024BarrereTSsim}. The former set-up is more relevant to model stellar radiative interiors, where the Tayler instability has been proposed to explain the angular momentum evolution of stellar interiors \citep{2019FullerTaylerSpruit}   
     and the magnetic dichotomy observed in intermediate-mass and massive stars  \citep{2010RudigerTS,2013SzklarskiApTS,2020BonannoTI,2020Jouve}.  
    On the other hand, a positive shear (with a faster-rotating surface) can be encountered in proto-neutron stars (PNSs). In this context, the TS dynamo plays a central role in a new scenario explaining magnetar formation from slow-rotating progenitors \citep{2022BarrereTS}.

    The TS has also been proposed as a mechanism for angular momentum transport potentially leading to their collapse into black holes \citep{2022MargalitAMomTransport}. This study estimated the angular momentum transport timescale due to the transport by the 
    saturated magnetic fields from analytical scalings but neglected the duration of the dynamo growth. 
    However, neutrino viscosity could make the TS dynamo stable, or it could require more time to reach saturation than the lifetime of the HMNS.    

    Therefore, this paper aims to study whether the conditions in BNS merger remnants could allow the TS dynamo to grow and impact its evolution. In particular, this study applies a similar one-zone model to that of \citet{2022BarrereTS}, within the framework of BNS mergers.
    The paper is organized as follows: in Sec.~\ref{sec:model}, we describe the modelling of the HMNS and the equations for a one-zone model of the TS dynamo. The results of the one-zone models are presented in Sec.~\ref{sec:Results}, followed by a comparison to the recent super-high resolution GRMHD simulation from \citet{2024Kiuchi} in Sec.~\ref{sec:Comp}. Finally, we discuss the validity of our assumptions in Sec.~\ref{sec:Disc} and conclude in Sec.~\ref{sec:Conclusion}.


\section{A one-zone model of the Tayler-Spruit dynamo in the hypermassive neutron star}
\label{sec:model}
\subsection{Modelling of the hypermassive neutron star}
\label{3Ddata}
\begin{figure}[ht]
   \centering
   \includegraphics[width=0.5\textwidth]{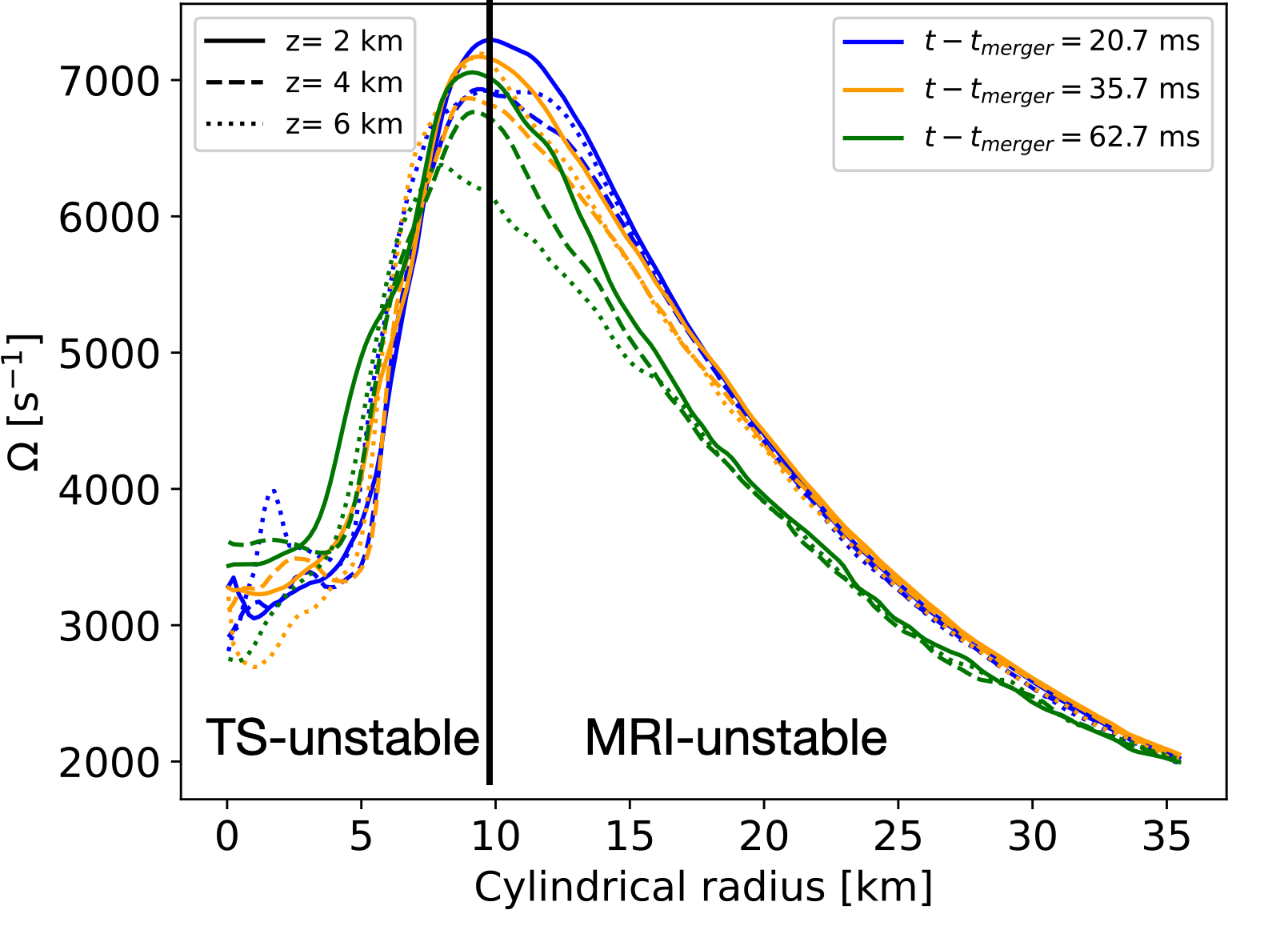}
   \includegraphics[width=0.5\textwidth]{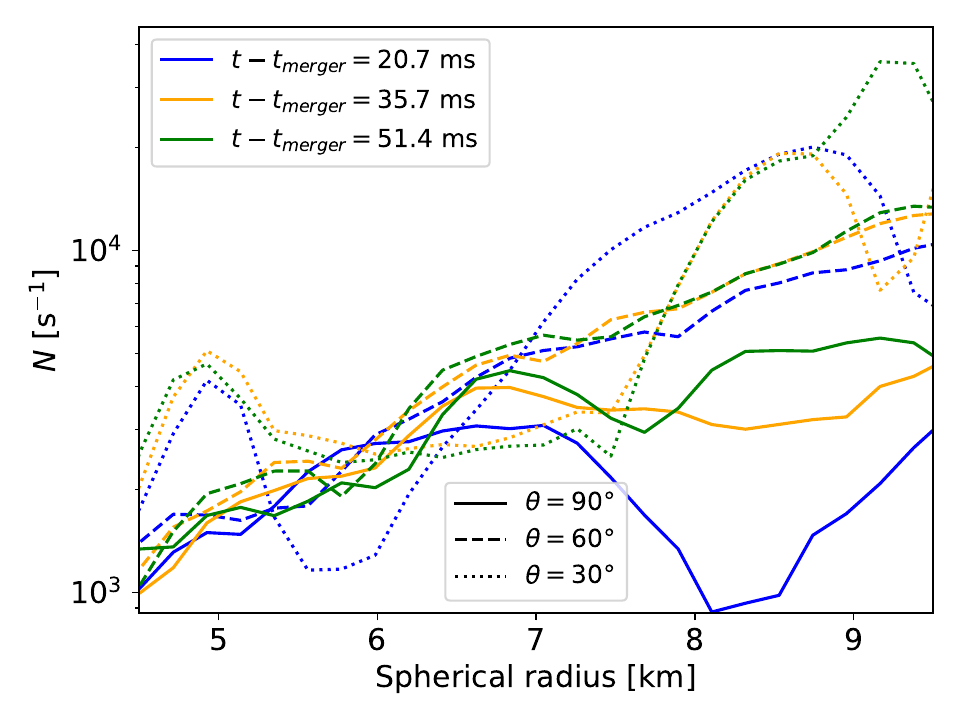}
   \caption{(Top) Azimuthal averaged rotation profile of the remnant HMNS from the 3D GRMHD simulation as a function of the cylindrical radius, at different times $t$ and vertical position $z$.
   (Bottom) Averaged Brunt-Vaisala frequency of the remnant in the TS-unstable region as a function of the cylindrical radius, at different times $t$ and different $\theta$ angle.}
   \label{fig:rot_profile}%
\end{figure}

For our HMNS model, we use the data from a 3D GRMHD simulation with a stiff equation of state, DD2~\citep{2010HempelDD2}, where the remnant lasts for more than $0.15$~s \citep{2024Kiuchi} (See Sec.~\ref{sec:Comp} for more detail).
To assess whether the TS dynamo is relevant for this HMNS model, we look at the rotation profile of the HMNS at different times (see Top panel of Fig.~\ref{fig:rot_profile}). 
The remnant displays a cylindrical rotation profile with a large region in positive differential rotation 
between a cylindrical radius of $R=5$ km and $R=10$ km\footnote{In the following, the cylindrical radius is denoted by $R$ and the spherical one by $r$.}, 
where the TS dynamo could grow.
We stress that this rotation profile is 
typical of HMNSs in many different simulations \citep{2006ShibataTaniguchiBNS,2016KastaunBNSstable,2017KastaunHMNSstable}. 
To characterise the differential rotation, we use the shear rate $$q \equiv \frac{d \ln \Omega}{d \ln R},$$ which means that a Keplerian shear corresponds to $q=-1.5$.
In the following, we use the cylindrical radius $R=7 $ km as a typical radius for our thermodynamical quantities. At this radius, we have $\Omega=5468 \rm \ s^{-1}$ and a shear rate of $q=1.12$.

Since the TS dynamo requires a stable stratification, we test whether the core of the BNS remnant is convectively stable \citep{2023RadiceBNS}.
For that, we use the relativistic Ledoux criterion, given by
\begin{equation}
    C_\mathrm{L} = \frac{\partial \rho(1+\epsilon)}{\partial r} - \frac{1}{c_s^2}\frac{\partial P}{\partial r},
\end{equation}
where $\rho, \ \epsilon, \ c_s$ and $P$ are the density, specific internal energy, sound speed and pressure, respectively. We find that the HMNS is indeed stably stratified, and we use the general relativistic formula for the Brunt-\Vaisala{} (BV) frequency $N$ derived in \citet{2013MullerBuoyancyGR}:
\begin{equation}
    N^2\equiv - c^2 \frac{\alpha C_\mathrm{L}}{\rho h \phi^4} \frac{\partial \alpha}{\partial r}\,, 
\end{equation}
where $\alpha, \ h$ and $\phi$ are the lapse function, specific enthalpy and conformal factor for the spatial metric: $N$ is presented in the bottom panel of Fig \ref{fig:rot_profile}.
This formula comes from the hydrostatic equilibrium written as $\mathcal{G} = - c^2 \nabla \ln \alpha$, where, in the Newtonian limit, $\mathcal{G}$ is the gravitational acceleration. 
Since our HMNS is rapidly rotating, we prefer to use this definition of hydrostatic equilibrium rather than the one based on
the pressure gradient for our background.
With this formula, we find typical values of $N=\SI{4970}{s^{-1}}$ at a radius of $R_{\rm TI}=\SI{7}{km}$ for the Brunt-\Vaisala{} frequency. The density is equal to $\rho = 3.7 \times 10^{14} \rm \ g \ cm^3$.

In this context, we cannot assume that the rotational angular velocity $\Omega$ is much lower than the Brunt-\Vaisala{} frequency. We also assume that the Alfv\'{e}n frequency remains small compared to the rotation and Brunt-\Vaisala{} frequencies,
\begin{equation}
    \omega_{\rm A} \equiv \frac{B}{\sqrt{4\pi \rho r^2}} \ll \Omega \,\,\mathrm{and}\, \,\omega_\mathrm{A} \ll N 
    \implies B \ll \SI{2e17}{G} \,,
\end{equation} 
for the one-zone model.
Note that this hypothesis may no longer be valid when we analyze the core dynamics in the GRMHD simulation, as the toroidal magnetic field can be winded to higher values. 
When these assumptions hold, the growth rate of the Tayler instability is $\sigma_{\rm TI}=\omega_{\rm A}^2/\Omega$. In the case of slow rotation $\Omega \ll \omega_A$, the growth rate is $\sigma_{\rm TI}=\omega_{\rm A}$.

\subsection{Newtonian Tayler-Spruit picture}

As in \citet{2022BarrereTS}, we decompose the velocity and magnetic field components into axisymmetric components, denoted as $B_i$ or $v_i$ and non-axisymmetric components, denoted as $\delta B_i$ and $\delta v_i$. We justify the Newtonian approach in the Appendix.

The shear rate evolution due to angular momentum transport by Maxwell stresses is written as
\begin{equation}\label{eq:q_evol}
    \partial_t {q} =-\gamma_{\rm AM}q= \frac{T^{MAX}_{R\phi}}{4 \pi \rho \Omega R_{\rm TI}^2} = -\frac{B_R B_{\phi}}{4\pi\rho\Omega R_{\rm TI}^2}\,,
\end{equation}
where $T^{MAX}_{ij}$ is the Maxwell stress tensor.
For the evolution of the angular velocity, 
we consider that the initial value decreases due to the magnetic torque. We also express it in terms of the Maxwell stress tensor, which gives
\begin{equation}\label{eq:omega_evol}
    \partial_t{\Omega}= \frac{R_{\rm TI}^{3} T^{MAX}_{R\phi}}{ I} =-\frac{R_{\rm TI}^{3} B_R B_{\phi}}{ I} 
    \,,
\end{equation}
where $I = 1.667\times 10^{45} \rm g \ cm^2$ is the moment of inertia of the inner zone below \SI{10}{km}. 
This means that in the case of a strong magnetic field, the rotation rate could decrease fast and be in the regime $\Omega \ll \omega_A$. In the following, we keep the growth rate of the Tayler instability $\sigma_{\rm TI}$ that is equal to
\begin{equation}
        \sigma_{\rm TI} = 
    \begin{cases}
            \omega_A & \text{if} \quad \Omega \ll \omega_A\,, \\
      {\omega_A^2}/{\Omega} &  \text{if}\quad  \omega_A \ll \Omega\,,\\
    \end{cases}
\end{equation}
in the evolution equations (Eqs.~\eqref{eq:Bperp_evol_sys1},~\eqref{eq:Bphi_evol1} and ~\eqref{eq:Br_evol_new1}) of the magnetic field. 
For the dissipation of the perturbed magnetic field, we consider the same damping rate $\gamma_{\rm cas} \sim \displaystyle\frac{\delta v_A}{r}$ 
as \citet{2019FullerTaylerSpruit}, where $\delta v_A \equiv \displaystyle\frac{\delta B_\phi}{\sqrt{4 \pi \rho}}$ is the Alfven velocity of the non-axisymmetric $\delta B_\phi$. 
It comes from the cascade rate of the perturbed magnetic field towards smaller scales.
The evolution equation of $\delta B_{\phi}$ is therefore
\begin{equation}\label{eq:Bperp_evol_sys1}
    \partial_t\delta B_{\phi}=\left(\mathcal{B}_{\rm TI} \sigma_{\rm TI}-\gamma_{\rm cas}\right)\delta B_{\phi}= \mathcal{B}_{\rm TI} \sigma_{\rm TI}\delta B_{\phi}- \frac{\delta v_{\rm A}}{R_{\rm TI}}\delta B_{\phi}\,,
\end{equation}
where $\mathcal{B}_{\rm TI}$ is a boolean that describes whether the HMNS is unstable to the Tayler instability (see Eq.~\eqref{eq:Tayler_bool}). In the case of ideal MHD, the instability is stabilised when the magnetic tension due to 
the radial magnetic field, $\sim k_{\rm TI} B_R \delta B$, where $k_{\rm TI}$ is the wavenumber of the Tayler instability, overcomes the driving force of the Tayler instability, the magnetic tension due to the toroidal magnetic field $\sim m/r B_\phi \delta B$, with $m=1$ being the main mode of the Tayler instability \citep{2019FullerTaylerSpruit,2024Skoutnev}. We have therefore
\begin{equation}
    \mathcal{B}_{\rm TI} = 
    \begin{cases}
            0 & \text{if}\quad B_R/B_\phi > \omega_{\rm A}/N,\\
      1 &  \text{otherwise}.\\
    \end{cases}
    \label{eq:Tayler_bool1}
\end{equation}

One other difference between the HMNS model and the PNS model of \citet{2022BarrereTS} is the geometry: the rotation profile in an HMNS is cylindrical while it is spherical for a late stage of the PNS with fallback accretion. In our study, we, therefore, consider the cylindrical radial field $B_R$ instead of the spherical radial field. 
The derivation of the equations is almost the same 
as it is just adapted to cylindrical coordinates. With a cylindrical differential rotation, the growth of the toroidal field $B_\phi$ is due to the winding of the cylindrical radial field $B_R$, described by $\partial_t B_\phi|_\mathrm{winding}  = q \Omega B_R$. The growth rate of the toroidal field is therefore $\sigma_{\rm shear} = q \Omega {B_R}/{B_\phi}$. As in \citet{2022BarrereTS}, we estimate the dissipation rates of $B_R$ and $B_\phi$ by the non-linear magnetic energy dissipation rate~$\dot{E}_{\rm mag}$ following 
\begin{equation}
    \gamma_{\rm diss} \equiv \frac{\dot{E}_{\rm mag}}{B_\phi^2},
\end{equation}
where the radial magnetic field is neglected compared to the dominant azimuthal field such that we can estimate $\dot{E}_{\rm mag} \sim B_\phi \partial_t B_\phi$. 
The change in axisymmetric magnetic energy is due to the growth of the Tayler instability, and thus, we have
\begin{equation}
    \dot{E}_{\rm mag} \sim \sigma_{\rm TI} |\delta B_\phi|^2.
\end{equation}
It finally gives a dissipation rate 
\begin{equation}
  \gamma_{\rm diss} \sim \sigma_{\rm TI} \left(\frac{\delta B_{\phi}}{B_{\phi}}\right)^2 
  \,.
\end{equation}
We take the simplification in order of magnitude of $\delta B_\perp \sim \delta B_\phi$, as $\delta B_\perp$ is not well defined for cylindrical coordinates.
The evolution equation of $B_{\phi}$ is therefore
\begin{equation}\label{eq:Bphi_evol1}
    \partial_t B_{\phi}=\left(\sigma_{\rm shear}-\gamma_{\rm diss}\right)B_{\phi}=q\Omega B_R-\sigma_{\rm TI}\frac{\delta B_{\phi}^2}{B_{\phi}}\,.
\end{equation}

In cylindrical coordinates, the growth rate $\sigma_{\rm NL}$ of $B_R$ comes from non-linear effects of the perturbed magnetic field and velocity. This corresponds to the toroidal component of the electromotive force $\mathcal{E}_\phi$ that is given by 
\begin{equation}
    \frac{\partial B_R}{\partial t} = - \frac{\partial \mathcal{E}_\phi}{\partial z} \sim \frac{\mathcal{E}_\phi}{L_z}, 
\end{equation}
where $L_z$ is the vertical wavelength of the Tayler instability. 
By definition, in cylindrical coordinates, the toroidal component of the electromotive force is given by $\mathcal{E}_\phi= \delta v_R \delta B_z - \delta v_z \delta B_R$. As in \citet{2022BarrereTS}, we assume $\mathcal{E}_\phi \sim \delta v_R \delta B_z \sim \delta v_R \delta B_\perp \sim \delta v_R \delta B_\phi$. By assuming 
a magnetostrophic balance between the Lorentz and Coriolis forces,
which gives $\delta v_\perp \Omega \sim \delta v_{\rm A} \omega_{\rm A}$, 
we retrieve the same formula as previous studies in spherical geometry if we assume $L_z \sim R_{\rm TI}$, which we will do for simplicity.
For consistency, we also assume that the non-linear growth
rate of the axisymmetric radial field is zero when the HMNS
is stable to the Tayler instability. The evolution equation of $B_R$ is therefore 
\begin{equation}
\label{eq:Br_evol_new1}
\partial_t B_R=\left(\sigma_{\rm NL}-\gamma_{\rm diss}\right)B_{R}=\mathcal{B}_{\rm TI} \frac{\omega_{\rm A}^2}{N\Omega}\frac{\delta B_{\phi}^2}{\sqrt{4\pi\rho R_{\rm TI}^2}}-\sigma_{\rm TI}\left(\frac{\delta B_{\phi}}{B_{\phi}}\right)^2 B_R
\,.    
\end{equation}
Overall, the magnetic field evolution is governed by equations similar to the equations of \citet{2022BarrereTS}, with some changes to take into account the growth rate and the stabilisation of the instability.

\subsection{Impact of diffusive processes on the Tayler-Spruit dynamo}

In an HMNS, the impact of 
neutrinos on the momentum equation can be modelled as a strong viscosity for scales larger than the neutrino mean free path. By taking into account the degeneracy of neutrinos in the core of the HMNS, the kinematic shear viscosity can be estimated by the formula \citep{1996KeilNeutrino,2015GuiletVisc}%
\begin{equation}
    \nu = 1.2 \times 10^{8} \left(\frac{\rho}{10^{14} \text{\ g cm}^{-3}}\right)^{-2} \left(\frac{T}{10 \rm MeV}\right)^{2} = 7.36 \times 10^7 \rm cm^2 \ s^{-1},
\end{equation}
at $R=\SI{7}{km}$. 
This would lead to a magnetic Prandtl number of 
\begin{equation}
    Pm \equiv \frac{\nu}{\eta} = 5.8\times 10^{12}, 
\end{equation}
where $\eta$ is assumed to be the resistivity due to electron-proton scattering \citep{1993ThompsonPNS}.
The thermal diffusion coefficient $\kappa$ due to neutrinos in core-collapse supernovae is higher than that by the viscous effect and the thermal Prandtl number $Pr$ is found to be 
\begin{equation}
   Pr \equiv \frac{\nu}{\kappa} = 10^{-3}\rm{-}10^{-2},
\end{equation}
\citep{2020ScienceRaynaud,2022ReboulSalze}, and we assume a similar ratio here.

We checked that the assumption of a magnetostrophic balance in the formalism of \citet{2019FullerTaylerSpruit} is verified and not changed to a viscous-Lorentz force balance with the neutrino viscosity (see Appendix \ref{app:force_balance}). 

The HMNS could be stabilised due to the fast dissipation of the perturbed velocity. We therefore estimate the critical magnetic field $B_{\phi,\rm crit}$ to be unstable by checking the allowed wavenumber $k_{\rm TI}$  of the instability. 
The minimum of this wavenumber has been estimated by \citet{2002spruitTS} as 
\begin{equation}
    k_{\rm TI} > \frac{N}{R_{\rm TI} \omega_{\rm A}}.
    \label{eq:k_lower}
\end{equation}
We also estimate that the effect of viscosity is not dominant if the growth timescale of the Tayler instability 
is faster than the viscous timescale, which limits the maximum wavenumber to be smaller than
\begin{equation}
    k_{\rm TI}^2 < \frac{\sigma_{\rm TI}}{\nu}= \frac{\omega_{\rm A}^2}{\Omega \nu}.
    \label{eq:k_upper}
\end{equation}
In order to satisfy both limits, the magnetic field must be stronger than the critical magnetic field  
\begin{equation}
    {B_{\phi,{\rm crit}}} = \left(4\pi \rho R_{\rm TI} N \right)^{\frac{1}{2}} \left(\nu \Omega \right)^{\frac{1}{4}} \sim 3.2 \times 10^{15} \text{G},
    \label{B_crit_classic}
\end{equation}
This value is much higher than that found in \citet{2022BarrereTS} because we consider the neutrino viscosity, which is much higher than the fluid viscosity or resistivity. 
Note that in the case of the slow-rotating limit, the growth rate changes and the critical magnetic field would be given by
\begin{equation}
    {B_{\phi,{\rm crit}}} = \left(4\pi \rho R_{\rm TI}\right)^{\frac{1}{2}} \left(\frac{\nu N^2}{R_{TI}^2} \right)^{\frac{1}{3}}. 
\end{equation}
For simplification, the wavelength of the Tayler instability $k_{\rm TI}$ is assumed to be equal to $k_{\rm TI, crit}$ for all the evolution of the one-zone model. 
This critical magnetic field strength gives the corresponding critical wavenumber 
\begin{equation}
    k_{\rm TI,crit}= \left(\frac{\sigma_{\rm TI,crit}}{\nu}\right)^{1/2}.
    \label{eq:k_TI_crit}
\end{equation} 
This high critical strength for our model means that the HMNS will be stable to the Tayler instability during the time needed for the toroidal field to be winded up by the shear.
We note that this critical magnetic field strength is a conservative estimate as the recent study of \citet{2024Skoutnev} shows that, in the case of a large cylindrical gradient of the toroidal field (corresponding to the geometric criterion of \citealt{1980GoossensTI}), the criterion depending on diffusivities is 
\begin{equation}
    \frac{\omega_A}{2\Omega} \gg \left(\frac{N}{2\Omega}\right)^{1/2} \left(\frac{\eta k_\theta^2}{2\Omega} \right)^{1/4} \min\left(1,\left(\frac{\eta}{\kappa}\right)^{1/4} \right) \approx 2.15 \times 10^{-9},
\end{equation}
where $k_\theta$ is the wavelength along the  vector $e_{\theta}$ in spherical coordinates.
This gives a lower limit for the magnetic field of $B_{\phi,\rm crit} > 1.1 \times 10^9$ G. This limit is easily met through the winding of the magnetic field, and hence, we keep the viscous criterion to be conservative.

In order to take into account the impact of the viscosity in the evolution equations, we modify the boolean $\mathcal{B}_{TI}$ (Eq.~\eqref{eq:Tayler_bool1}) in the evolution equation of $\delta B_{\phi}$ (Eq.~\eqref{eq:Bperp_evol_sys1}) 
to include the cases where the HMNS is stable to the Tayler instability. It gives now 
\begin{equation}
    \mathcal{B}_{\rm TI} = 
    \begin{cases}
            0 & \text{if}\quad B_\phi< B_{\phi,\rm crit} \quad\text{or}\quad B_R/B_\phi > \omega_{\rm A}/N,\\
      1 &  \text{otherwise}.\\
    \end{cases}
    \label{eq:Tayler_bool}
\end{equation}

The system can therefore be summarized  by the following equations 
\begin{equation}\label{eq:Bphi_evol}
    \partial_t B_{\phi}=\left(\sigma_{\rm shear}-\gamma_{\rm diss}\right)B_{\phi}=q\Omega B_R-\sigma_{\rm TI}\frac{\delta B_{\phi}^2}{B_{\phi}}\,,
\end{equation}
\begin{equation}\label{eq:Bperp_evol_sys}
    \partial_t\delta B_{\phi}=\left(\mathcal{B}_{\rm TI} \sigma_{\rm TI}-\gamma_{\rm cas}\right)\delta B_{\phi}= \mathcal{B}_{\rm TI} \sigma_{\rm TI}\delta B_{\phi}- \frac{\delta v_{\rm A}}{R_{\rm TI}}\delta B_{\phi}\,,
\end{equation}
\begin{equation}
\label{eq:Br_evol_new}
\partial_t B_R=\left(\sigma_{\rm NL}-\gamma_{\rm diss}\right)B_{R}=\mathcal{B}_{\rm TI} \frac{\omega_{\rm A}^2}{N\Omega}\frac{\delta B_{\phi}^2}{\sqrt{4\pi\rho R_{\rm TI}^2}}-\sigma_{\rm TI}\left(\frac{\delta B_{\phi}}{B_{\phi}}\right)^2 B_R
\,.    
\end{equation}

\section{Results of the one-zone model}
\label{sec:Results}

\subsection{A fiducial case}

\begin{figure}[ht]
   \centering
   \includegraphics[width=0.5\textwidth]{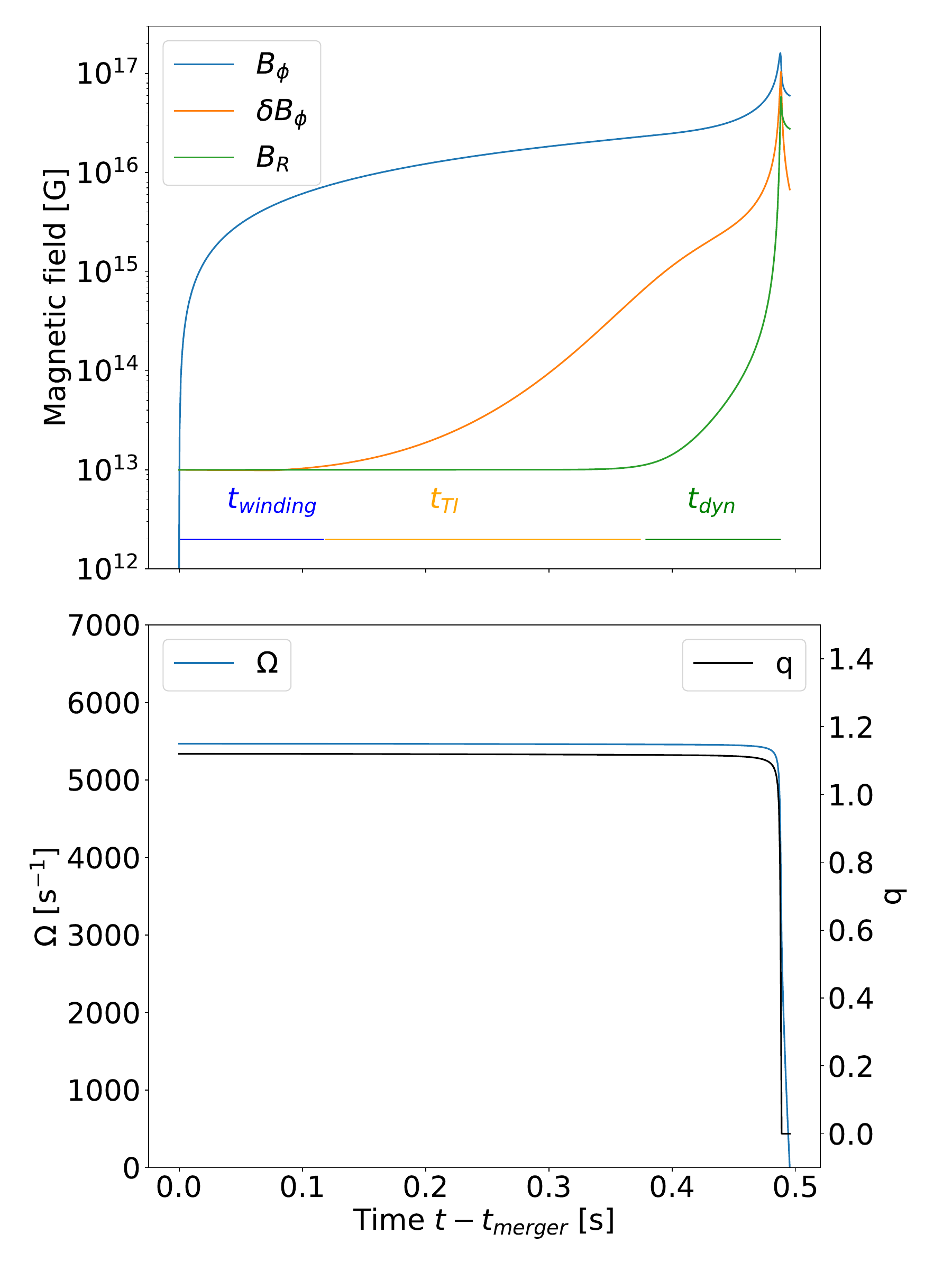}
   \caption{\textit{Top:} time evolution of the magnetic field strength for a typical HMNS with thermodynamic quantities taken at $t-t_\text{merger}= \SI{0.03}{s}$ at a radius of $R_{\rm TI} = \SI{7}{km}$ and with an initial magnetic field strength of $B_{R,0} = 10^{13}$\,G. \textit{Bottom:} time evolution of the rotational angular velocity $\Omega$ and the shear rate $q$.}
    \label{fig:B13_evol}%
\end{figure}

We first consider a fiducial case, in which 
the initial large-scale radial magnetic field is assumed to be $B_{R,0} = 10^{13}$ G. The perturbed magnetic field also needs to be initialised, and we consider that the first perturbations have a similar order of magnitude, $\delta B_{\phi,0} = B_{R,0} = \SI{e13}{G}$. 
The azimuthal magnetic field is initialised to $B_{\phi,0} = 1$\,G 
but this choice has little influence, as the winding is very efficient at amplifying this component of the magnetic field. 
Figure \ref{fig:B13_evol} shows the time evolution of the magnetic field, which divides into three phases as expected: 
\begin{enumerate}
    \item[(i)] First, the shear amplifies the toroidal magnetic field. This winding phase lasts for $\simeq 0.15$~s. 
    
    \item[(ii)]  When the toroidal magnetic field is stronger than the critical magnetic field $B_{\phi, \rm crit}$, the non-axisymmetric magnetic field $\delta B_\phi$ grows due to the Tayler instability. The duration of this phase is $\simeq0.25$~s.
    
    \item[(iii)] When $\delta B_\phi$ becomes strong enough for the non-linear induction to be relevant, the dynamo sets in
    and increases all the magnetic field components for the next 0.08~s. The dynamo stops when the angular momentum is transported and the HMNS core settles in a rigid rotation with a low angular velocity (see bottom panel in Fig.~\ref{fig:B13_evol}).
\end{enumerate}

At saturation, the toroidal, radial and perturbed magnetic fields are equal to
\begin{align}
    B_\phi^{\rm sat} &= \SI{1.6e17}{G} \label{eq:Bp_sat} \,,\\ 
    B_R^{\rm sat} &= \SI{5.9e16}{G} \label{eq:Br_sat} \,,\\
    \delta B_\phi^{\rm sat} &= \SI{1.0e17}{G} \,.
\end{align}
With the saturated toroidal magnetic field, we have $\displaystyle\frac{\omega_{A,sat}}{\Omega} = 0.58$, meaning that the initial assumption of $\omega_A \ll \Omega$ is verified for most of the dynamo phase until the last rapid growth where the rotational angular velocity decreases to low values due to the magnetic torque.
We compare the saturation values of the axisymmetric field $B_R$ and $B_\phi$ obtained by the model to the estimated saturated magnetic field from the following formulas \citep{2019FullerTaylerSpruit}
\begin{align}
\label{eq:Bp_fuller}
B_{\phi}^{\rm F, sat} &= \sqrt{4\pi \rho R_{\rm TI}^2} \Omega \left(\frac{q\Omega}{N}\right)^{1/3}  \approx 2.8 \times 10^{17} \ \rm G , \\ 
\label{eq:Br_fuller}
B_r^{\rm F, sat} &=  \sqrt{4\pi \rho R_{\rm TI}^2} \Omega \left(\frac{q^2\Omega^5}{N^5}\right)^{1/3} \approx 3.3\times 10^{17} \ \rm G,\\
\label{eq:dBp_fuller}
\delta B_\phi^{\rm F,sat} &= \sqrt{4\pi \rho R_{\rm TI}^2}\Omega \left(\frac{q\Omega}{N}\right)^{2/3}  \approx 3.0\times 10^{17} \ \rm G.
\end{align}

These values differ from the results of our models because some of the model's assumptions are not valid before reaching the predicted saturated values.  Indeed, both Fuller's and Spruit's predictions give $\omega_A > \Omega$ and $\omega_A > N$ with our values. These values are not reached, as the differential rotation is quenched before. Nonetheless, the saturated values still agree within a factor of $3$--$5$.

It is interesting to note that the alternative model of \cite{2002spruitTS} predicts similar magnetic field strength at saturation:
\begin{gather}
\label{eq:Bp_spruit}
B_\phi^{S,sat} = \sqrt{4\pi \rho R_{\rm TI}^2} q \left(\frac{\Omega^2}{N} \right)  \approx 3.2\times 10^{17} \ \rm G\\
B_r^{S,sat} = \sqrt{4\pi \rho R_{\rm TI}^2}  q^2 \left(\frac{\Omega^4}{N^3}\right) \approx  3.9\times 10^{17} \ \rm G .
\end{gather}
This can be interpreted by the fact that the rotation and Brunt-Väsälä frequencies are of the same order of magnitude.

With this initial magnetic field, the TS-dynamo could generate an intense dipole and an even stronger toroidal magnetic field when the long-lived HMNS survives for longer than 0.5\,s after the BNS mergers.

\subsection{Dependence on the initial dipole magnetic field strength}

\begin{figure}[t]
   \centering
\includegraphics[width=0.49\textwidth]{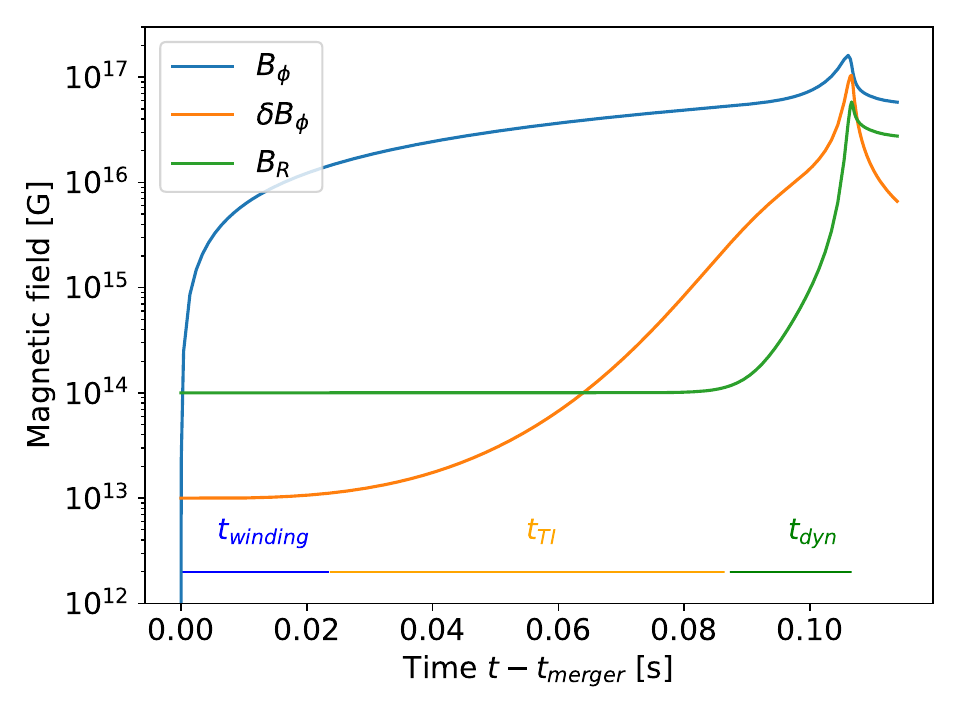} \\
   \includegraphics[width=0.49\textwidth]{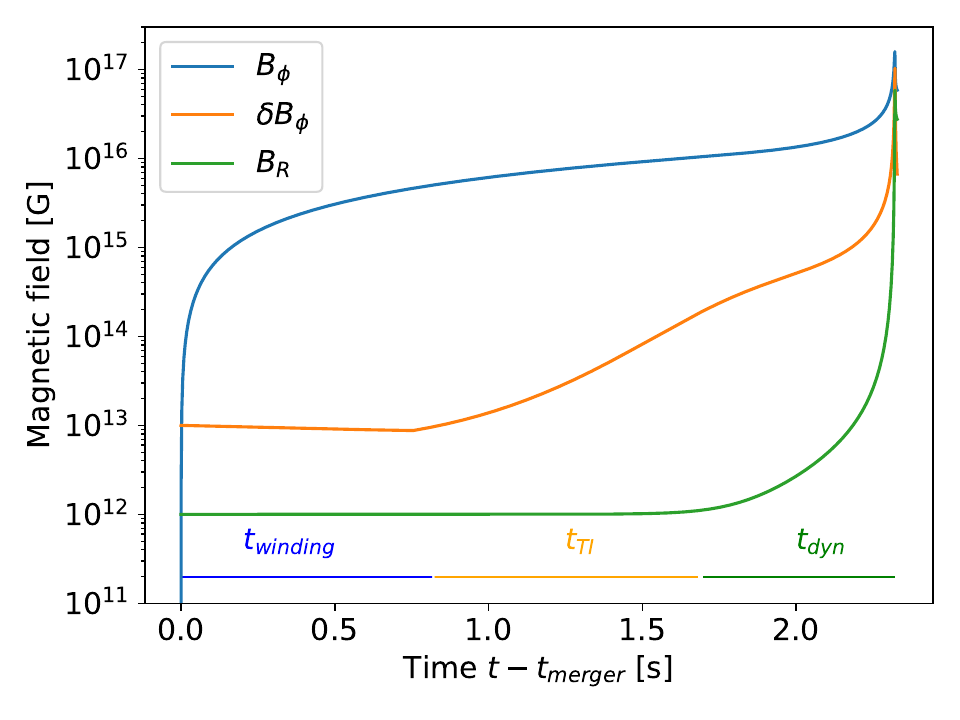} \\
\includegraphics[width=0.49\textwidth]{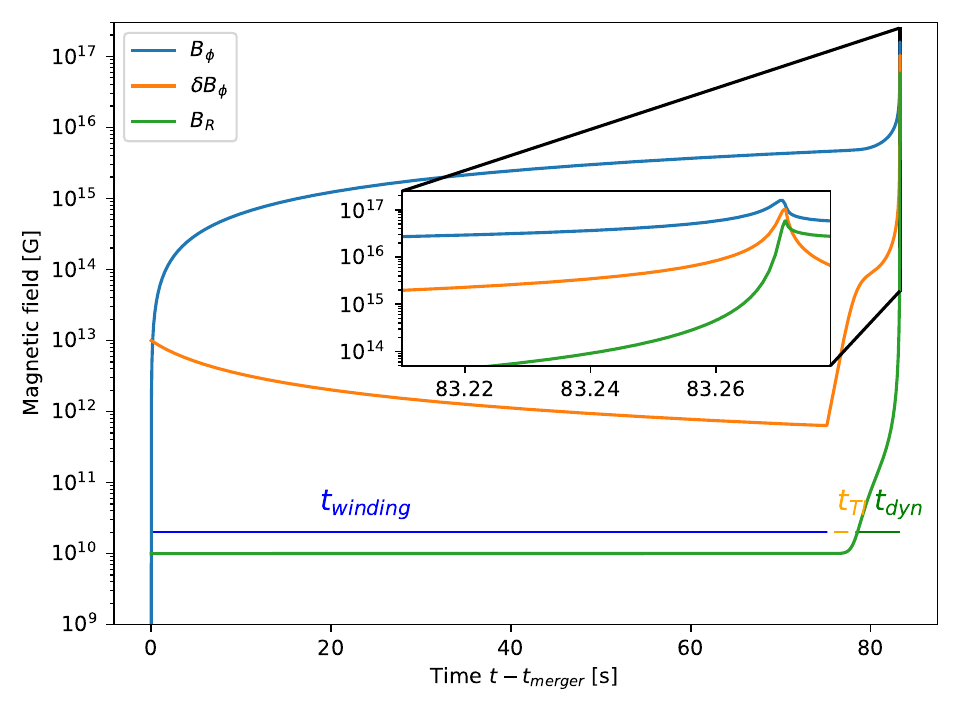}
   \caption{Time evolution of the magnetic field  for different initial magnetic dipole strength $B_{R,0}$ with $B_{R,0}=10^{14}$ G (\textit{top}), $B_{R,0}=10^{12}$ G (\textit{middle}) and $B_{R,0}=10^{10}$ G (\textit{bottom}).} 
    \label{fig:comp_init_b}%
\end{figure}

For BNS mergers, the duration of the dynamo is quite essential as it may determine the time at which 
the remnant collapses into a black hole. 
This duration is expected to depend on the initial intensity of the large-scale poloidal magnetic field, which is uncertain as discussed above.
We therefore vary the initial magnetic field strength to study its influence on the dynamo. Figure \ref{fig:comp_init_b} shows the time evolution for different initial magnetic fields $B_{R,0}$. While the maximum magnetic field strength is similar for all cases, the time to reach saturation $\Delta t_{\rm sat}$ decreases drastically with the increase of the initial magnetic field strength (see Tab.~\ref{tab:times}).
\begin{table}[t]
    \centering
    \caption{Initial magnetic field strength $B_{R,0}$ and time to reach saturation in the one-zone model. 
    }
    \begin{tabular}{cl}
        \toprule
     $B_{R,0} \;\si{[G]}$ & $ \Delta t_{\rm sat} \;\si{[s]} $ \\
     \midrule
        $ 10^{14} $ & $\approx 0.11 $\\
        $ 10^{13} $ & $\approx 0.48$ \\
        $ 10^{12} $ & $\approx 2.4 $\\
        $ 10^{10} $ & $\approx 86 $ \\ 
        \bottomrule 
    \end{tabular}
    \label{tab:times}
\end{table}

To realistically estimate the initial dipole field strength, the magnetic field in the core of the HMNS, amplified by the Kelvin-Helmholtz (KH) instability at the onset of the merger, should be explored with dissipation \citep{1999RasioReviewKH,2014KiuchiKH,2018KiuchiGRMHD,2024Kiuchi}. 
Due to this instability, the large-scale initial magnetic field could be amplified 
and potentially be stronger than $B_{R,0} =10^{13}$ G. Therefore, the TS dynamo could saturate in less than \SI{0.5}{s}. 

To complete this picture, we can estimate the duration of each phase of the dynamo process.
The first phase is the winding phase, during which the toroidal field is amplified. It lasts until either the toroidal field becomes strong enough for the Tayler instability to grow as fast as $B_\phi$ \citep{2022BarrereTS}, or that it becomes strong enough to reach the critical magnetic field for the instability to start with the high viscosity. The formula for the winding time is therefore 
\begin{equation}
    t_{\rm winding} = \frac{\max ( \omega_{\rm A,TI}, \omega_{\rm A,crit})}{q\Omega \omega_{R,0}} = \max \left((q^2\omega_{R,0}^2\Omega)^{-1/3} , \frac{\omega_{\rm A,crit}}{q\Omega \omega_{R,0}}\right),
\end{equation}
where $\omega_{\rm A,crit}$ and $\omega_{R,0}$ are the Alfv\'{e}n frequency of the critical magnetic field due to viscosity and the initial Alfv\'{e}n frequency, respectively. $\omega_{\rm A,TI}$ is the Alfv\'{e}n frequency when the Tayler instability becomes dynamically relevant, i.e. when
$\sigma_{\rm shear}= \sigma_{\rm TI}$, which is given by 
\begin{equation}
    \omega_{\rm A,TI} = \left (q \Omega^2 \omega_{R,0} \right)^{1/3}.
\end{equation}
At the end of the winding phase, the toroidal magnetic field is equal to 
\begin{equation}
    B_\phi(t_{\rm winding}) =   q  \Omega \ t_{\rm winding} B_{R,0}
    \,.
    \label{eq:Bphi_winding}
\end{equation}

For the Tayler instability phase, we adapt Eq.~(51) of  \cite{2022BarrereTS} considering that $B_\phi(t_{\rm winding})$ is given by Eq.~\eqref{eq:Bphi_winding}. Introducing the corresponding Alfv\'{e}n frequency $\omega_{A, \rm winding}$, we find
\begin{equation}
    t_{\rm TI} \sim \frac{\Omega}{(\omega_{A, \rm winding})^2} \ln \left(\frac{\omega_{A, \rm winding}^2r}{\Omega \ \delta v_{A,0}}\right),
\end{equation}
where $\delta v_{A,0}=\delta v_{A}(t=0)$.

Finally, we use the same formula as \citet{2022BarrereTS} for the dynamo phase
\begin{equation} 
    t_{\rm dyn}= \left(\frac{B_\phi}{\partial_t^2 B_\phi}\right)^{1/2} = \frac{\Omega}{\omega_{A,\rm dyn}^2} \left(\frac{N}{q \omega_{A,\rm dyn}}\right)^{1/2},
\end{equation}
where $\omega_{A,\rm dyn} = \left(q N \Omega^4 \omega_{R,0}^2\right)^{1/7}$ is the Aflv\'en frequency for which the winding rate is equal to the non-linear growth rate $\sigma_{\rm NL}$ of the radial magnetic field $B_r$.

\begin{figure}[t]
   \centering
    \includegraphics[width=0.5\textwidth]{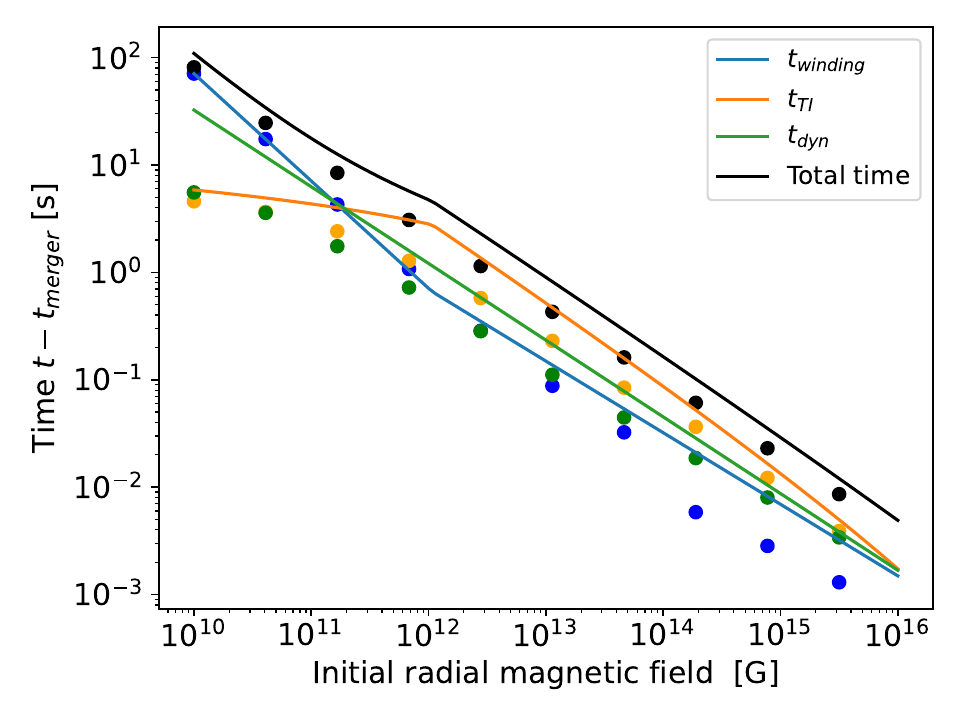}
   \caption{Timescale of the different phases post-merger of the TS dynamo as a function of the initial radial magnetic field strength. The points are measured from the one-zone model, while the lines represent analytical estimates. 
   }
              \label{fig:timescale}
\end{figure}

Figure \ref{fig:timescale} compares these timescale estimates to the ones measured in the one-zone model. The analytical predictions are overall in good agreement with only a slight overestimation compared to the model results.
We also give a simple fit of the total dynamo time 
\begin{equation}
 t_\mathrm{total} \approx 65.7 \left(\frac{B_{R,0}}{10^{10}}\right)^{-0.71} \rm s \label{fit_time}.   
\end{equation}
This shows that the TS dynamo can grow in a few seconds as long as the initial magnetic field strength is higher than \SI{e12}{G}.
This figure shows that the shear time dominates the total time for initial magnetic field strength lower than \SI{e12}{G}. In this case, a higher rotational angular velocity or shear rate would linearly decrease the total dynamo time. 
For higher initial magnetic field strength, the Tayler instability phase dominates the total time and therefore the Tayler-instability growth rate becomes a key parameter. 

\section{Comparison with a 3D GRMHD simulation}
\label{sec:Comp}

In the previous section, we showed that the TS dynamo is able to grow and reach saturation on a timescale of a few tens of milliseconds for an initial dipole higher than $10^{15}$ G, which is the typical dipole value used in GRMHD simulations of BNS mergers \citep[e.g.,][]{2013GiacomazzoSMNS, 2014KiuchiKH,2020MostaMagnetarKilo,2024Kiuchi}. %
As a consequence, the magnetic fields would be expected to grow from the TS dynamo in such GRMHD simulations. 
In this section, we analyze the data of a 3D ideal GRMHD simulation in which the massive neutron star remnant is long-lived \citep{2024Kiuchi}. {Since the remnant is differentially rotating during the duration of the simulation (see Figure~\ref{fig:rot_profile}), we call it an HMNS in the following. Note that it may become a supramassive neutron star once in solid body rotation or collapse to a black hole in the case of a strong spin-down. 

Since we look for the TS dynamo, we focus our analysis on the core of the BNS merger remnant at radii lower than $10$ km. In this region, the resolution is $\Delta x = 12.5$ m in the range $[0,4.5]$ km and $\Delta x = 25$ m in the range $[4.5,9]$ km,  
which is roughly the region where the positive differential rotation is found. In this region, the resolution becomes $\Delta x = 100$ m after $t - t_\mathrm{merger}= 0.050$\,s.}
The numerical techniques and the study of the other regions of these simulations are detailed in \citet{2024Kiuchi}. Note that the outer surface of the merger remnant core for which $q$ is negative is subject to the $\alpha\Omega$ dynamo as reported in \citet{2024Kiuchi}.

\subsection{Dynamics of the GRMHD simulation}

\begin{figure}[ht]
   \centering

    \includegraphics[width=0.5\textwidth]{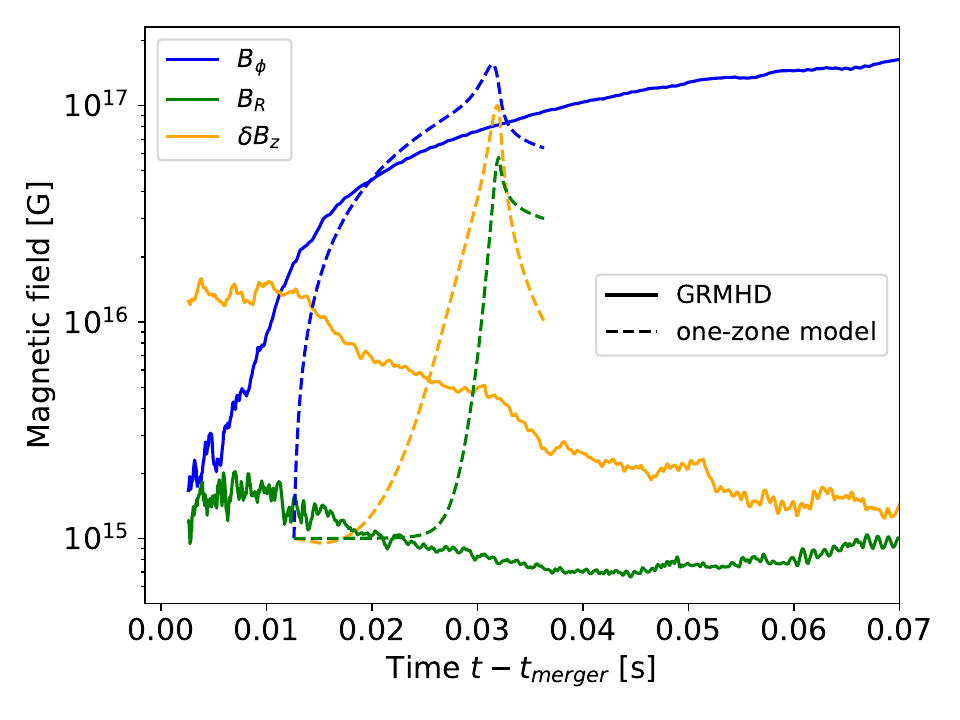}
   \caption{Time evolution of the axisymmetric magnetic field and $\delta B_z$ for the GRMHD simulation (solid lines) in the unstable region (see Section~\ref{3Ddata} and Figure \ref{fig:Geo_crit}) and for the one-zone model (dashed lines) that has been shifted temporally. 
   }
              \label{fig:comp_models}%
\end{figure}

The 3D neutrino radiation GRMHD simulation for a BNS merger of a symmetric binary case, with masses of 1.35-1.35 $M_{\odot}$ and a stiff equation of state, DD2 \citep{2010HempelDD2}, shows that the merger results in a HMNS that survives for timescales $> 0.15$~s. The rotation profile in the equatorial plane of the HMNS shows a differential rotation with $q > 0$ until $10$ km (see Fig.~\ref{fig:rot_profile}). 
This leads to the winding of the poloidal magnetic field in the core of the remnant.
Figure \ref{fig:comp_models} compares the time evolution of the magnetic field strength in the 3D GRMHD simulation and the one-zone model. 
The radial magnetic field in the one-zone model is amplified to its saturation values within 0.02~s, while in the GRMHD simulation, the radial magnetic field strength varies slightly 
and overall remains constant, staying close to its initial value around $10^{15}$ G.  
This shows that no TS dynamo processes occur in the simulation's HMNS core even though it would be expected according to the theoretical prediction. In addition, the non-axisymmetric vertical field $\delta B_z$ is only decreasing in this region, while it should be growing to a similar amplitude as the $\delta B_\phi$ of the one-zone model. 

\begin{figure}[ht]
   \centering
    \includegraphics[width=0.45\textwidth]{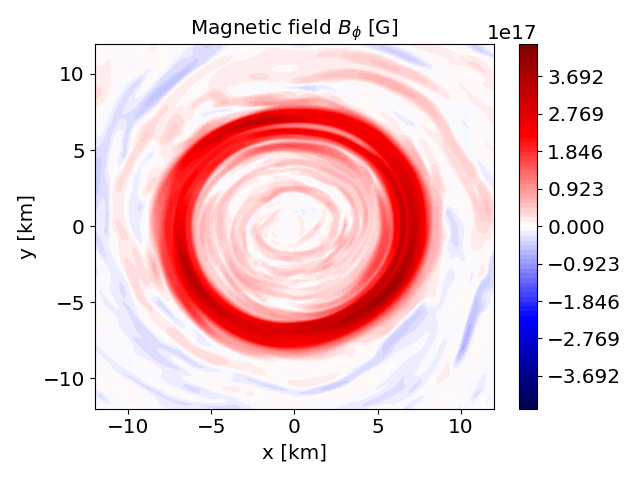} \\
     \includegraphics[width=0.45\textwidth]{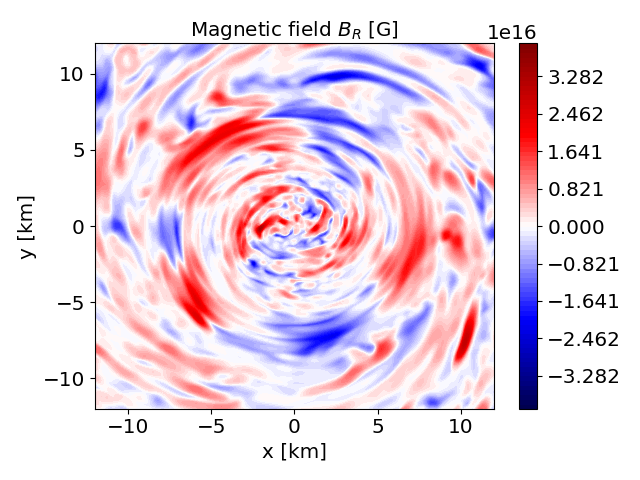}   \\   
    \includegraphics[width=0.45\textwidth]{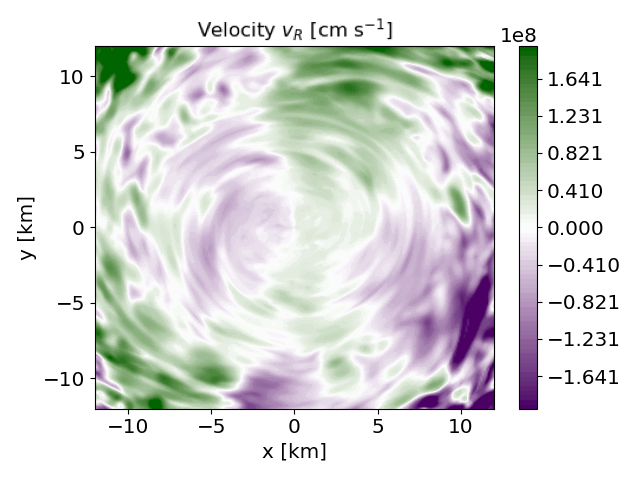}
   \caption{Snapshot of the toroidal magnetic field (\textit{top}), radial magnetic field (\textit{middle}) and radial velocity (\textit{bottom}) in the core of the HMNS, taken in the plane $z=\SI{2}{km}$ at $t -t_{\rm merger} = \SI{0.035}{s}$.} 
    \label{fig:state30ms}
\end{figure}

To understand why there is no TS dynamo, we look at a snapshot of the simulation in a plane at $z=2$ km and $t - t_{\rm merger}=\SI{0.03}{s}$ (see Fig.~\ref{fig:state30ms}). We see that the toroidal magnetic field is amplified to a maximum value of $4\times10^{17}$ G (top panel of Figure \ref{fig:state30ms}). Since this is larger than the critical magnetic field for a strong viscosity, 
the simulation should be unstable to the Tayler instability but
the non-axisymmetric component of the radial magnetic field does not show any evidence for the Tayler instability. Indeed, its main mode is an $m=2$ mode, and the geometry corresponds well to the velocity field in the equatorial plane (bottom panel of Fig.~\ref{fig:state30ms}). 
The origin of this $m=2$ mode is well-known and comes from the initial geometry and mass ratio of the binary. 
Figure \ref{fig:state30ms} shows that the velocity field is not a purely $m=2$ mode as the initial $m=2$ mode can also degrade as a $m=1$ mode, but it is different from the $m=1$ mode from the Tayler instability. 
Indeed, the $m=2$ and degraded $m=1$  modes give the radial magnetic field exactly the same geometry as the radial velocity field and it has the ring shape of the toroidal field. 
In addition, according to simulations of the Tayler instability \citep{2023JiTaylerinstability,2024BarrereTSsim}, we would expect to see some opposing polarities of the radial field on a meridional slice, contrary to the geometry observed in Figure~\ref{fig:Geo_crit} which is the same as the toroidal field. 

Positive differential rotation can stabilize the Tayler instability, as shown by \citet{2011KiuchiTayler}, 
but the toroidal magnetic field is decreasing vertically in the north hemisphere in the unstable region (see Fig.~\ref{fig:Geo_crit}), which makes it unstable (see Eq.~(27) of \citealt{2011KiuchiTayler}).
From this criterion, the core of the remnant at $t - t_{\rm merger} = \SI{0.06}{s}$ should be unstable to the Tayler instability with a fast-growing rate of $\sigma_{\rm TI} \sim \omega_A$, because we are in the "slow" rotating regime. 

To be unstable to the Tayler instability, the geometry of the magnetic field is also important, as the size of the magnetic field can limit the wavelength.
To estimate what is the maximal wavelength available, we used the geometrical instability criterion for a wave from \citet{1980GoossensTI}
\begin{equation}
    \frac{B_\phi^2}{4\pi R_{\rm TI}^2 \sin \theta} \left (m^2 - 2 \cos^2 \theta -2 \sin \theta \cos \theta \, \partial_\theta \log B_\phi  \right) < 0
    \,.
    \label{eq:geo_crit}
\end{equation}
Figure \ref{fig:Geo_crit} shows the toroidal magnetic field with the grey contours showing where the Tayler instability is stable according to this criterion with $m=1$. 
The magnetic field should indeed 
be unstable to the Tayler instability, but the unstable domain is relatively small, 
with a typical size of around $\SI{1}{km}$. Contrary to the one-zone model, the hypothesis $\omega_A \ll \Omega$ is not verified in the unstable domain.

Lastly, another instability criterion is that
${B_R}/{B_\phi} < \displaystyle\frac{k_\phi}{k_R}$ 
which implies that the magnetic tension due to the radial component should not exceed the magnetic pressure due to the toroidal component (Eq. \eqref{eq:Tayler_bool1}) to operate the Tayler instability. This criterion is usually given for an axisymmetric $B_R$ but it is unclear whether the magnetic tension from a non-axisymmetric $B_{R,m\neq 0}$ would stabilize the instability by overcoming the magnetic tension from the background toroidal field. 
 
Looking at the middle panel of Fig.~\ref{fig:state30ms}, this criterion may explain why the Tayler instability is temporarily stabilised as the radial length of $B_R$ is quite small. However, after some time, $B_\phi$ grows in strength and size with the winding (Figure~\ref{fig:Geo_crit}). At the same time, the amplitude of the non-axisymmetric modes decreases and the instability criterion becomes verified. 

In any case, these non-axisymmetric modes amplify the magnetic field to a lower strength than the ones in the one-zone model and are expected to dissipate less than a hundred milliseconds after the merger as the gravitational wave luminosity decreases \citep{2023RadiceBNS}. These modes might cause a delay in the appearance of the TS dynamo, but this does not explain why it does not appear later in the simulation.

\subsection{Numerical dissipation analysis}

The absence of the Tayler instability in this 3D GRMHD simulation could be due to the wavelength of the Tayler instability being lower than the grid resolution or to the fact that the numerical resistivity/viscosity reduces the growth rate of the instability.
Thus, we estimate the expected wavelength of  the Tayler instability if it were growing in the simulation. In ideal MHD, the wavelength would be limited by the buoyancy and would give $\lambda_{\rm TI}= \displaystyle\frac{\omega_A}{N} R \approx R$ in the simulation. This wavelength is well resolved so we need to look at the impact of diffusion on the wavelength.
From the instability criterion, the wavelength $\lambda_{\rm crit}=2\pi/k_{\rm TI, crit}$ below which the Tayler instability should occur is  
\begin{equation}
    \lambda_{\rm crit} = \frac{2 \pi R_{\rm TI} \omega_{\rm A,crit}}{N} = 
    \begin{cases}
    \displaystyle 2 \pi R_{\rm TI} \left(\frac{\Omega}{N} \right)^{1/2}  \left(\frac{\nu}{R_{\rm TI}^2 \Omega}\right)^{1/4} & \text{if}\quad \omega_A \ll \Omega, \\ \\
   \displaystyle  2 \pi R_{\rm TI} \left(\frac{\nu}{R_{\rm TI}^2 N}\right)^{1/3} & \text{if}\quad \Omega \ll \omega_A. 
    \end{cases}
    \label{lambdaTI}
\end{equation} 

As this wavelength depends on the numerical resistivity/viscosity inside the simulation, 
we estimate the numerical resistivity/viscosity. 
We use the fitting formula derived for Eulerian MHD codes in \citet{2017RembiaszNumericalVisc}
\begin{equation}
    \nu_\star = \mathcal{R}^{\Delta x}_\nu \mathcal{V} \mathcal{L} \left ( \frac{\Delta x}{\mathcal{L}}\right)^r, \\.
    \label{num_visc}
\end{equation}
where $\mathcal{V}$ and $ \mathcal{L}$ are the characteristic speed and length of the system, $\mathcal{R}^{\Delta x}_\nu$ and $r$ are fitting parameters.
We use the fitting parameters $r=4.95$ and $\mathcal{R}^{\Delta x}_\nu=42$ for the HLLD solver with the fourth-order Runge Kutta method, as it is the same Riemann solver used in our GRMHD simulations. We use the fast magnetosonic speed for the characteristic velocity as found in the study $\mathcal{V}= \max (v_A, c_s) \approx c_s \approx 1.2 \times 10^{10}$ cm s$^{-1}$ for the characteristic speed. We take the characteristic length as a variable $\mathcal{L} =\lambda_{\rm crit}$. 
Since the numerical resistivity depends on the resolution and on the scale we are considering, by combining Eqs.~\eqref{lambdaTI} and \eqref{num_visc}, we have to solve the equation
\begin{equation}
    \begin{cases}
          \lambda_{\rm crit}^{1.9875} = 2 \pi R_{\rm TI} \left(\frac{\Omega}{N} \right)^{1/2}  \left(\frac{\mathcal{R}^{\Delta x}_\nu \mathcal{V} \Delta x^{4.95} }{R_{\rm TI}^2 \Omega}\right)^{1/4},  & \text{if}\quad \omega_A \ll \Omega, \\ \\
          \lambda_{\rm crit}^{2.317} = 2 \pi \left(\frac{\mathcal{R}^{\Delta x}_\nu \mathcal{V} \Delta x^{4.95} R_{\rm TI} }{ N}\right)^{1/3},  & \text{if}\quad \omega_A \gg \Omega,
    \end{cases}
\end{equation}
which, with $\Delta x = \SI{100}{m}$, gives $\lambda_{\rm crit} = \SI{1.55}{km}$. Note that, in the fast-rotating limit ($\omega_A \ll \Omega$), we get $\lambda_{\rm crit}= \SI{2.4}{km}$ at a $R_{\rm TI}= \SI{7}{km}$. 
In the case where the geometry does not constrain the wavelength, the Tayler instability would be resolved by a hundred-meter resolution.

\begin{figure}[ht]
   \centering
    \includegraphics[width=0.5\textwidth]{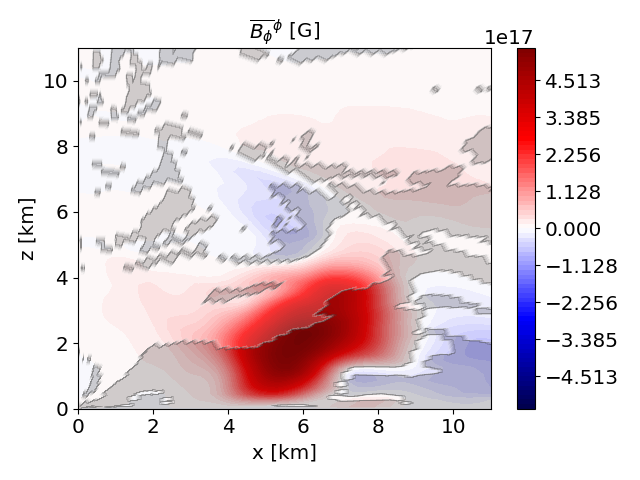}
    \includegraphics[width=0.5\textwidth]{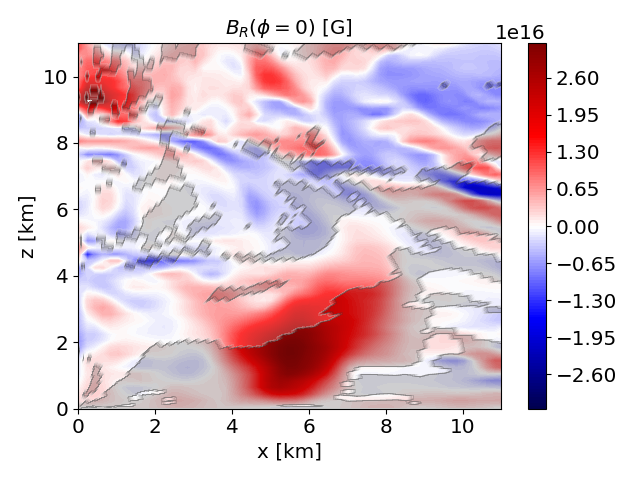}
   \caption{Snapshot of the axisymmetric toroidal magnetic field (top) and a slice of the radial magnetic field $B_R$ 
   (bottom) in the core of the remnant neutron star at $t -t_{\rm merger} = 0.060$ s and $y= 0$ km. The grey contours correspond to the regions stable to the Tayler instability according to the geometric criterion (Eq.~\eqref{eq:geo_crit} with $m=1$). 
   }
    \label{fig:Geo_crit}
\end{figure}

However, the unstable region of the magnetic field is limited by its vertical length $\approx \SI{1}{km}$ (Figure~\ref{fig:Geo_crit}), and we therefore have 
to limit the wavelength of the Tayler instability. 
To check the impact of dissipation on a mode with a 
$1$ km wavelength, we solve the local dispersion relation of \citet{1978Acheson} with equal diffusivities,
$\nu=\eta=\kappa$. We also assume the magnetic field to be increasing with radius and decreasing vertically, and we take the other quantities the same as the simulation (see Appendix \ref{app:Linear} for more details). 
Figure \ref{fig:growth_rate} shows that the growth rate in the limit of low dissipation is in fair agreement with $\sigma_{\rm TI} \approx \omega_A$, which is consistent with the regime $\omega_A \gg \Omega$. 
The stability limit due to dissipation is found to be in good agreement with the criterion $\omega_A = k_{\rm TI}^2 \eta$ (dashed line in Fig.~\ref{fig:growth_rate}). We also note that even for a high dissipation like the neutrino viscosity $\nu_{\rm HMNS} = 7.36 \times 10^7 \rm \ cm^2 \ s^{-1}$, the growth rate is unchanged compared to the ideal MHD case. We would therefore expect the instability to grow in the absence of numerical limitations. 

\begin{figure}[ht]
   \centering    \includegraphics[width=0.49\textwidth]{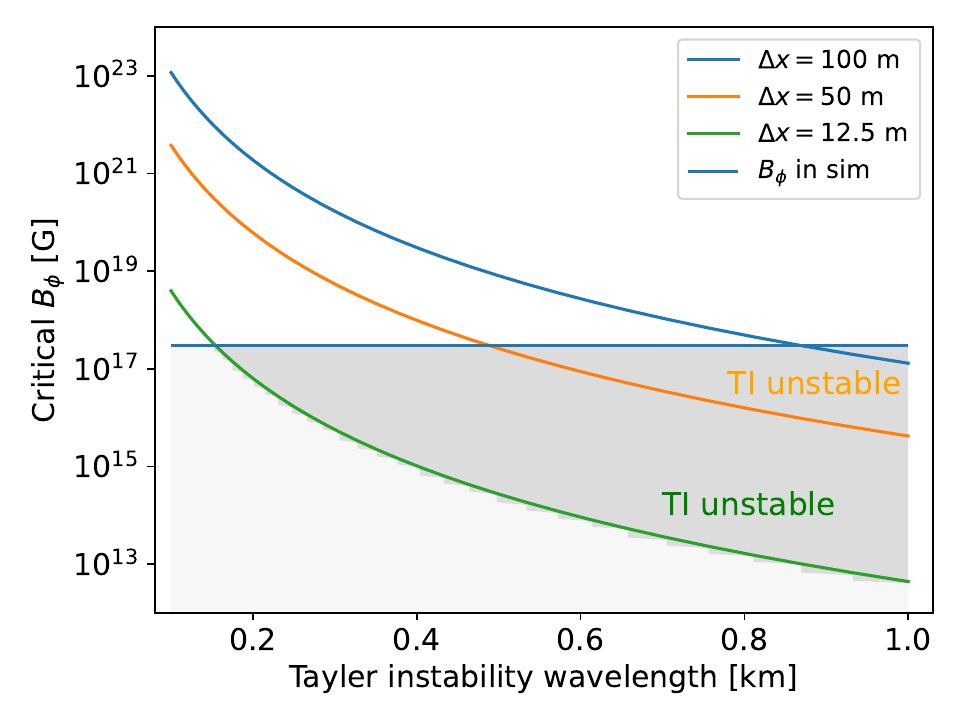}
   \caption{Critical magnetic field $B_{\phi, \rm crit}$ (Eq.~\eqref{B_crit_classic}) in the numerical simulation depending on the wavelength of the Tayler instability for several resolutions with the numerical viscosity and resistivity estimated with the simulation grid size (Eq.~\eqref{num_visc}). The Tayler instability should be unstable in the simulation when the critical field is weaker than the magnetic field in the GRMHD simulation (grey region for $\Delta x =12.5$ m).} 
    \label{fig:critical_Bphi}%
\end{figure}

We thus use the limit given by Eq.
\eqref{B_crit_classic} with the numerical dissipation as the criterion to see whether the Tayler instability can develop in the numerical simulation.
Figure \ref{fig:critical_Bphi} shows the critical magnetic field for the numerical viscosity/resistivity in the code depending on the length scale below \SI{1}{km} for different resolutions. 
The magnetic field found in the simulation is close to the critical magnetic field with the estimated dissipation of a 100-meter resolution, 
which may explain why the simulation is stable to Tayler instability at this resolution. 

\begin{figure}[ht]
   \centering
    \includegraphics[width=0.5\textwidth]{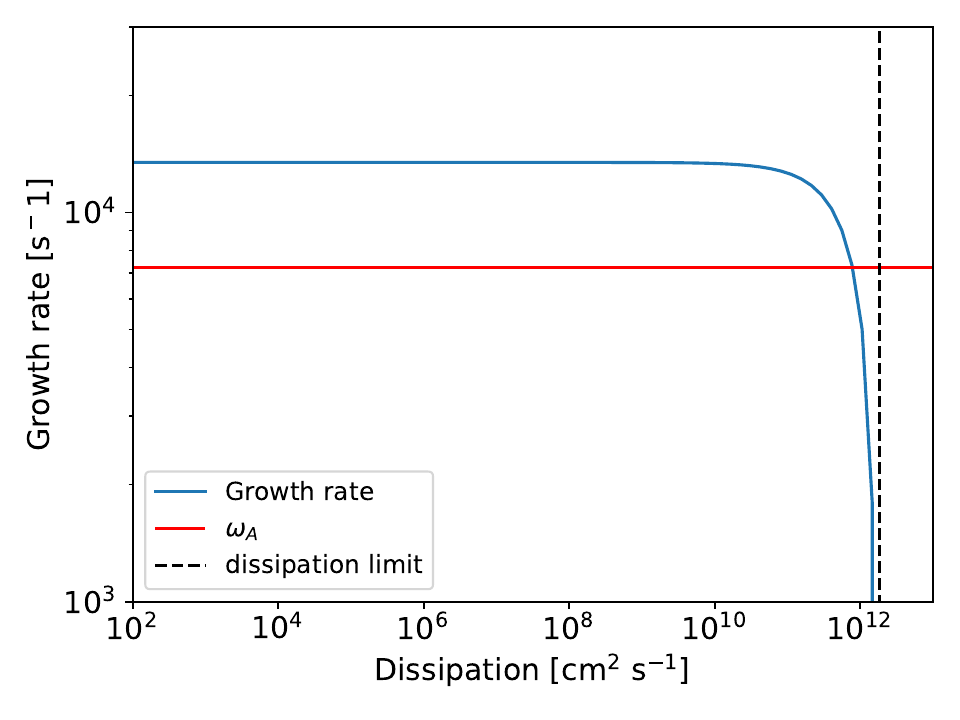}
   \caption{Growth rate of the Tayler instability as a function of dissipation, with $\nu=\eta=\kappa$. The red line shows the theoretical growth rate in the regime of $\Omega \ll \omega_A$. The dashed line is the stabilizing limit given by $k_{\rm TI}^2 \eta = \omega_{\rm A}$. 
   }
              \label{fig:growth_rate}%
\end{figure}

Overall, in our understanding, the Tayler instability may be stabilised by the non-axisymmetric modes initially and  
later on by the resolution that is not enough to resolve the small region where it should be unstable.

\section{Discussion}
\label{sec:Disc}

\subsection{Comparison with Boussinesq simulations}
\label{sec:Boussinesq}
Like the one-zone model of~\cite{2022BarrereTS}, the model of this article relies on strong assumptions to overcome the non-linearity of the dynamo mechanism, especially for the saturation phase~\citep{2002spruitTS,2007ZahnTS,2019FullerTaylerSpruit}. A comparison with the recent 3D numerical simulations of the TS dynamo with a positive differential rotation~\citep{2023BarrereTSsim,2024BarrereTSsim} is therefore useful to test the validity of these necessary assumptions. Their simulations show the existence of several TS dynamo branches, but we will only consider the strong, dipolar branch, which has been thoroughly studied by~\citet{2024BarrereTSsim} and harbours the strongest magnetic fields. Furthermore, this solution is in good agreement with the scaling laws in Eqs.~\eqref{eq:Bp_fuller}--\eqref{eq:dBp_fuller}, derived by~\citet{2019FullerTaylerSpruit}, which correspond to the formalism used in our one-zone model, with, however, a normalisation factor significantly lower than 1.

In the region of the HMNS core with a positive shear ($\SI{5}{}$--$\SI{10}{km}$), the stratification is in the range $N/\Omega \approx 0.5$--$1$, which is covered by the investigation of~\citet{2024BarrereTSsim}. By using the parameters of a typical HMNS ($R=\SI{7}{km}$, $\rho=\SI{3.7e14}{g.cm^{-3}}$, and $\Omega=\SI{5468}{s^{-1}}$) and extrapolating to $q\sim1$, we infer from the numerical models of~\citet{2024BarrereTSsim} with $N/\Omega=0.5$ and $N/\Omega = 1$ that the axisymmetric toroidal, poloidal magnetic fields and the magnetic dipole should reach
\begin{align}
    B^{m=0}_{\rm tor}&\approx 0.63\rm{-}\SI{1.4e16}{G}\,,\\
    B^{m=0}_{\rm pol}&\approx 0.15\rm{-}\SI{1.5e15}{G}\,,\\
    B_{\rm dip}&\approx 0.79\rm{-}\SI{6.3e14}{G}\,,
\end{align}
respectively. $B^{m=0}_{\rm tor}$ is therefore weaker by a factor of $\sim 20$--$45$ than $B_{\phi}^{\rm F,sat}$ (Eq.~\eqref{eq:Bp_fuller}). Likewise, $B^{m=0}_{\rm tor}$ and $B_{\rm dip}$ are $\sim 220$--$4200$ times weaker than $B_r^{\rm F, sat}$ (Eq.~\eqref{eq:Br_fuller}), which is consistent with the fact that the TS dynamo in the simulations of~\citet{2024BarrereTSsim} produces weaker large-scale magnetic fields than predicted by the formalism of~\citet{2019FullerTaylerSpruit}. Note, however, that these simulations are likely to underestimate the magnetic field strength in a realistic HMNS. Indeed, for computational reasons, the simulations were run with much higher resistivities than realistic estimates, which hinders the dynamo process. Moreover, the quantities $B^{m=0}_{\rm tor}$, $B^{m=0}_{\rm pol}$, and $B_{\rm dip}$ are volume-averaged. This suggests that the magnetic fields must be stronger locally when they are concentrated in some regions of the integration domain, which happens when the stratification increases.

\subsection{MRI-driven dynamo vs TS dynamo}

As it is argued in \cite{2022MargalitAMomTransport}, the TS dynamo could also be relevant for the transport of angular momentum where the differential rotation is decreasing with radius. However, depending on the Brunt-\Vaisala{} frequency, this region would be unstable to the MRI, which has a much faster growth rate than the TS dynamo in the fast-rotating case. For this reason, the MRI is expected to saturate first, and the Tayler instability would not impact this region.
To confirm this picture, we compute the growth rate of both instabilities 
in the equatorial plane at $r=\SI{20}{km}$ with a magnetic field $B_0 = 10^{15}$ G as it is found after the Kelvin-Helmholtz instability. We obtain
\begin{eqnarray}
   && \sigma_{\rm MRI}= \frac{q\Omega}{2} \approx \SI{2600}{s^{-1}} \nonumber \\ 
   && \sigma_{\rm TS} = \frac{\omega_{\rm A}^2}{\Omega} \approx 0.072 \left(\frac{B_0}{10^{15} \rm G}\right)^2 \rm \ {s^{-1}} \  \ll \sigma_{\rm MRI},  
\end{eqnarray}
due to $\omega_{\rm A} \ll \Omega$. 
Interactions from the turbulence developed by the MRI and the TS dynamo would occur at the radius where the differential rotation changes sign, making this transition region difficult to study.

\subsection{Angular momentum transport and spin-down time}

To have the complete picture of the angular momentum transport in the HMNS, we first describe the transport in the outer region, $R \gtrsim 10$\,km, which is unstable to the MRI. 
The angular momentum transport timescale due to MRI turbulence is 
\begin{equation}
    t_{\rm MRI} \approx \frac{R^2}{\alpha_{\rm turb} H \Omega} \sim 5 \times 10^{-3}  \left(\frac{\alpha_{\rm turb}}{0.1}\right)^{-1} \left(\frac{\Omega}{\Omega_K}\right)^{-1} \left(\frac{H}{R/3}\right)^{-2} \quad[\rm s ]. 
\end{equation}
The spin-down in the outer region would be from the magnetic torque of the large-scale magnetic field generated by the $\alpha\Omega$ dynamo. %
On the other hand, our study shows that in the inner core, the angular momentum transport timescale due to the TS dynamo corresponds to the time it takes to reach saturation. 
Indeed, the very large saturated magnetic fields lead to very fast transport of angular momentum on a much shorter timescale than the dynamo growth timescale. 
This growth ranges from $10^{-2}-10^{2}$ s and scales with the initial magnetic field as $t_{\rm total} = 65.7 \displaystyle\left(\frac{B_{R,0}}{10^{10}}\right)^{-0.71}$ s. 
One interesting feature is that due to the positive differential rotation, the angular momentum would be transported inwards rather than outwards.

The spin-down time depends on how we consider the generated magnetic field inside the remnant. We first consider that the magnetic field stays buried in the core of the HMNS, then it would act as a torque and slow the core down following Eq.~\eqref{eq:omega_evol}.
By taking the saturated magnetic field values of the one-zone model (Eq.~\ref{eq:Bp_sat}-\ref{eq:Br_sat}), the core would slow down at a rate of 
\begin{equation}
    \gamma_{\rm spin-down, core} \equiv \frac{\dot{\Omega}}{\Omega} = - \frac{ R_{\rm TI}^{3}
    B_R B_\phi}{I \Omega} \approx \SI{3.5e2}{s^{-1}} 
    \,.
\end{equation}
We can also assume that the obtained dipole becomes the dipole at the surface of a remnant neutron star without matter outside. Then, the dipole formula gives the following spin-down rate of
\begin{equation}
    \gamma_{\rm spin-down, dipole} \equiv \frac{\dot{\Omega}}{\Omega} = -\frac{B_r^2 R_{\rm TI}^6 \Omega^2}{6 c^3 I}   \approx 
    \SI{4.5e-2}{s^{-1}}
\end{equation}
assuming that the saturated value of the dipole is $B_r \sim B_R =4\times 10^{16}$ G.
The spin-down due to the magnetic torque would therefore be faster than the dipole spin-down and slow the remnant down 
on a timescale of milliseconds. 

Hence, the remnant would collapse in a few milliseconds when the dynamo saturates and the rotational kinetic energy would not be emitted in electromagnetic waves. 
If the magnetic braking follows the torque used in this study,
this could therefore shorten the lifetime of a remnant neutron star after the merger depending on the initial radial field $B_{R,0}$.
However, if we consider the saturated values from \citet{2024BarrereTSsim}, the spin-down value due to the magnetic torque would be a thousand times slower $\gamma_{\rm spin-down, core} \approx 0.33 \rm s^{-1}$. The collapse time would then be around O(1) s after the dynamo saturation rather than O(0.010) s. 

\section{Conclusions}
\label{sec:Conclusion}

We have investigated the TS dynamo in the context of BNS mergers. Following \cite{2022BarrereTS}, we developed a one-zone model of the TS dynamo for a BNS merger remnant with realistic parameters estimated 
from a 3D GRMHD simulation. 

We found that the TS dynamo could develop in the core of remnant neutron stars due to its positive differential rotation. Due to the impact of neutrino viscosity, the toroidal magnetic field must be amplified to higher values than $3.2 \times 10^{15}$ G 
by the winding in order to be Tayler-unstable.

The magnetic field's evolution can be divided into three phases: the winding phase, the Tayler instability phase and the non-linear growth of the dynamo. Saturation occurs when the differential rotation is effectively quenched by the Maxwell stress. The magnetic field generated by the dynamo saturates at a very high intensity of $B_\phi=1.6 \times 10^{17}$\,G and $B_R=5.9 \times 10^{16}$\,G for the magnetic dipole, according to the one-zone model.  
These strong intensities must be taken with caution as current simulations of the TS dynamo applied to PNSs predict a weaker toroidal field of $ 10^{16}$ G  and a poloidal field of $10^{15}$ G.

The saturated magnetic field strength does not depend on the initial magnetic field, unlike the time required to reach saturation,  
which can range from 0.1\,s for a magnetic dipole of \SI{e14}{G} to 2.4\,s for a magnetic dipole of \SI{e12}{G}, which becomes comparable to 
the $O(1)$\,s, i.e., HMNS lifetime in the long-lived case. In the first case, the transport of angular momentum by the TS dynamo would lead to a faster collapse to a black hole or a faster spin-down of the remnant. 
This shows that the TS dynamo would be important in the case of a long-lived remnant neutron star as long as the initial magnetic field dipole is higher than $10^{12}$ G. 

However, these results depend on whether the TS dynamo grows fast enough, and consequently on the resulting magnetic dipole after the Kelvin-Helmholtz instability. 
A study using realistic initial magnetic fields in remnant neutron stars and with a converged growth of the Kelvin-Helmholtz instability is therefore needed. 

As we discussed in section \ref{sec:Boussinesq}, the simulations by \citet{2024BarrereTSsim} give weaker saturated magnetic field strengths than our results but the differences could be due to a cylindrical rotation profile and the different background stratification. The value of the magnetic dipole needs to be confirmed by 3D numerical simulation with a rotation profile and a background stratification corresponding to a remnant neutron star. This is left to a further study,
which should also address the question of the interaction between the TS dynamo and the MRI instability that operates at cylindrical radii larger than \SI{10}{km}.

The TS dynamo operating in the remnant neutron star could have significant consequences for astrophysical observations. First, the angular momentum transport and the magnetic braking would happen when the dynamo saturates, which would lead to the collapse of the remnant neutron star if it is more massive than the maximum mass of a non-rotating neutron star $M_{\rm TOV}$. Otherwise, it would lead to a stable, slow-rotating neutron star with a strong magnetic field, a proto-magnetar. If the rotation is damped as fast as described by the magnetic torque of the saturated field, the lifetime of the remnant neutron star would be reduced and could prevent the late-time emission of a fast-rotating proto-magnetar. These results, in particular the rate of the magnetic braking, need to be tested with  
numerical simulations.

In addition, having motions due to the TS dynamo in the neutron star remnant is expected to lead to some emission in gravitational waves, which might be detectable for a close event. Using Fuller's formalism, we estimate the amplitude of the velocity perturbations due to TS dynamo with the formula 
\begin{equation}
    \delta v_{r,\rm sat} = \delta v_{A,\rm sat} \frac{\omega_{A,\rm sat}^2}{N \Omega} \approx 4.5 \times 10^9 \ \rm cm \ \rm \ s^{-1}\,.
\end{equation}
For the sake of simplicity, we estimate the strain by using the order of magnitude derived from the quadrupole formula \citep{2002Kokkotas}
\begin{equation}
    h \approx \frac{G}{c^4}\frac{\varepsilon E_{\rm turb}}{D_{GW}}
    \,,
\end{equation}
where $\varepsilon$ is the efficiency to convert the turbulent energy to gravitational waves that we assume to be low, with $\varepsilon \approx 5\%$. The kinetic energy is computed by assuming that this velocity is constant in the region from $5$ to $10$ km and oscillates at the Brunt-\Vaisala{} frequency $N$. A distance of \SI{100}{Mpc} would then give a strain of $h \approx \SI{2e-25}{}$, 
which is of the same order but slightly lower than the sensitivity of the future generation of gravitational detectors.
Knowing that the kinetic energy predicted here can be viewed as optimistic, a direct detection of the signal will be difficult but the TS dynamo would still impact the collapse time of the remnant.
Further investigations of the TS dynamo in BNS mergers are therefore important to better understand future multi-messenger observations.

\begin{acknowledgements}
    
    This work was supported by the European Research Council (MagBURST grant 715368), and the  PNPS and PNHE programs of CNRS/INSU, co-funded by CEA and CNES. 
    This work was in part supported by the Grant-in-Aid for Scientific Research (grant Nos. 23H04900, 23H01172, 23K25869) of Japan MEXT/JSPS.
\end{acknowledgements}

%
%
   \bibliographystyle{aa} 
   \bibliography{biblio} 

\appendix 

\section{Effects from general relativity}

In the context of BNS merger remnant neutron stars, effects from general relativity can become important and the formalism needs to be adapted to a 3+1 decomposition. The three-component velocity $v^i$, which corresponds to the Newtonian velocity in the non-relativistic limit, is defined by 
\begin{equation}
    v^i= \frac{u^i}{u^t},
\end{equation}

where $u^\mu$ is the four-component velocity, $u^t$, $u^j$ being its time and space components.
In our simulation, the shift vector can be neglected 
at radii larger than 
$5$ km. We can also assume that the movement of the flow is non-relativistic meaning that $u^t =  \frac{W}{\alpha} \approx \frac{1}{\alpha}$, where $W$ is the Lorentz factor of the flow. 

In ideal GRMHD, the induction equation is the same as the Newtonian case but for the magnetic field in the inertial frame $\mathcal{B}^i$, written as $B_i$ in the following, and the three-component velocity $v^i$. The magnetic field that enters the momentum equation is the magnetic field comoving with the fluid flow $b^\mu$ and the spatial components are defined as 
\begin{equation}
    b_i=\frac{1}{\alpha u^t}\left( B_i + u_i \left(B^j u_j \right)\right)
    \,.
\end{equation}
Knowing that $u_\phi \sim \frac{1}{\alpha} v_\phi$ would be dominant the component of the velocity and that $v_\phi$ is of the order of magnitude $0.1 c$, the order of magnitude of the corrections would be  $4\%$ for $b_\phi$ as it would scale as $u_\phi^2$ and lower for the other components. 
Another correction would be in the expression of the angular momentum transport by the Maxwell stress tensor which is then given by
\begin{equation}
    T^{MAX}_{R\phi} = b^2 u_R u_\phi - b_R b_\phi
    \,,
\end{equation}
where $b^2 = b^\mu b_\mu = \frac{1}{\alpha^2 (u^t)^2} \left(B^2 + (B^i u_i)^2\right)$. The corrections would be of the order of $\frac{b_\phi}{b_R} u_R u_\phi$ and therefore around $\sim 8\%$ as ${b_\phi}/{b_R}\sim 2$.

In GRMHD, the definition of the Alfv\'{e}n velocity also changes to 
\begin{equation}
    v_A^2 = \frac{b^2}{4\pi \rho h + b^2}
    \,,
\end{equation}
where $h$ is the specific enthalpy.
In the case of $b^2 \ll 4\pi \rho h$, the formalism can be simply adapted by replacing $\rho$ by $\rho h$. This assumption corresponds to $v_A \ll c$, which holds already in the case of $\omega_A \ll \Omega$ as the rotational velocity $R_{\rm TI} \Omega$ is lower than $c$. This correction would reduce the Alfv\'{e}n frequency and thus the growth rate $\sigma_{\rm TI}$ by $h(R_{\rm TI}) \sim 1.126 c^2$ in the case of fast rotation. 

\section{Force balance at high viscosiy}
\label{app:force_balance}
\citet{2019FullerTaylerSpruit} 
make the assumption of
a Coriolis-Lorentz force balance for the perturbed velocity. However, another balance of the viscous forces and Lorentz force could be possible. 
To compare the Coriolis force and the viscous force, a local Ekman number can be defined as 
\begin{equation}
    E = \frac{\nu k_{\rm TI}^2}{\Omega},
\end{equation}
where $k_{\rm TI}$ is the wavenumber of the displacements due to the Tayler instability. When $E<1$, the Coriolis force is stronger than the viscous forces and a magnetostrophic balance can 
be safely assumed.

The Ekman number based on the conservative critical magnetic field $B_{\phi, \rm crit}$ (Eq. \eqref{B_crit_classic}) and its wavelength (Eq. \eqref{eq:k_TI_crit}) is $E \approx 1.6 \times 10^{-6} \ll 1 $, so we can safely assume the magnetostrophic balance. 
From Eq.~\eqref{eq:k_lower}, 
we can estimate the range of magnetic fields where  viscosity will dominate over Coriolis force, i.e. $E > 1$, which is given by
\begin{gather}
    {B_\phi < B_{\phi,\text{min, visc}}} = \sqrt{4 \pi \rho} N \sqrt{\frac{\nu}{\Omega}} \approx \SI{2.9e13}{G}
    \,.
\end{gather}
Therefore, the magnetostrophic balance can be assumed for both the Tayler instability and the dynamo phases, but not necessarily 
for the initial stages of the time evolution.
However, the magnetic field strength is amplified to superior values than $B_{\text{min, visc}}$ by the winding, for which the magnetostrophic balance can also be assumed. 
Since $B_{\phi,\rm min, visc} < B_{\phi, \rm crit}$, the magnetostrophic balance can be assumed during the Tayler instability phase and dynamo phase. However, in the winding phase, $B_\phi$ could be smaller than $B_{\phi,\rm min, visc}$ depending on the initial value of $B_r$. In such a case, a viscous-Lorentz force balance should be assumed. However, as this phase is not very important for the growth of the perturbed magnetic field and, therefore, we can assume a magnetostrophic balance for the whole duration. 
The viscous-Lorentz balance can be written as $\delta v_\perp \sim \delta v_{\rm A} \omega_{\rm A} /(\nu k_{\rm TI}^2)$, which is derived from the magnetostrophic balance $\delta v_\perp \sim \delta v_{\rm A} \omega_{\rm A} /\Omega$ divided by the Ekman number. The magnetostrophic balance is used to derive the non-linear growth rate of the TS dynamo, and the balance needs to be divided by the Ekman number $E$ with the viscous-Lorentz. 
In case we wanted to adapt Eq.~\eqref{eq:Br_evol_new}, the growth rate of $B_R$ would then be divided by $E$ if $E>1$, which gives 
\begin{multline}
\label{eq:Br_evol_ek}
    \partial_t B_R=\left(\sigma_{\rm NL}-\gamma_{\rm diss}\right)B_{R}=\mathcal{B}_{\rm TI}\frac{\omega_{\rm A}^2}{N\Omega}\frac{\delta B_{\perp}^2}{\text{max}(1,E) \sqrt{4\pi\rho R_{\rm TI}^2}} \\
     -\frac{\omega_{\rm A}^2}{\Omega}\left(\frac{\delta B_{\perp}}{B_{\phi}}\right)^2 B_R\,.
\end{multline}

\section{Linear study of Tayler instability}
\label{app:Linear}
We solve the dispersion relation of \citet{1978Acheson} with Python. 
We first use symPy to calculate the polynomial coefficients from the relation and then solve the roots of the polynomial numerically. We tested that we recover the dispersion relation in both the non-rotating ideal case,  and the rotating case as in \citet{2011KiuchiTayler}. We can recover the theoretical growth rate in the two opposite limits $\omega_A \ll \Omega$ and $\Omega \ll \omega_A$. The code used to solve the linear dispersion relation is publicly available at the following address \url{https://github.com/alexisreboulsalze/AchesonLinSolve/tree/main}.

\end{document}